\documentclass[final,5p,times,twocolumn]{elsarticle}
 \biboptions{comma,sort&compress}
\usepackage{graphicx}
\usepackage{here}
\usepackage[dvips]{epsfig}
\usepackage{cuted}
\def\beq{\begin{equation}}
\def\eeq{\end{equation}}
\def\bea{\begin{eqnarray}}
\def\eea{\end{eqnarray}}

\def\nnb{\nonumber}
\def\ga{\left(}
\def\dr{\right)}

\def\lrar{\Longrightarrow}
																
\def\nnb{\nonumber}

\def\la{\langle}
\def\ra{\rangle}

\def\ba{\vspace*{-0.2cm}\begin{array}}
\def\ea{\end{array}\vspace*{-0.2cm}}

\def\b{$\bullet~$}

\def\als{\alpha_s}

\def\gg2{ \la\alpha_s G^2 \ra}
\def\gg3{g^3f_{abc}\la G^aG^bG^c \ra}
\def\ggg4{\la\als^2G^4\ra}

\journal{Elsevier}

\begin{document}
%\markboth{Stephan Narison, Montpellier (FR)}{ }
\begin{frontmatter}
%%%%%%%%%%%%%%%%%%%%%%%%%%%%%%%%%%%%%%%%%%%%%%%%%
%\begin{document}
\title{Spectra and Decay Constants of $B_c$-like and $B^*_0$ Mesons in QCD} 
 \author[label2]{Stephan Narison}
\address[label2]{Laboratoire
Particules et Univers de Montpellier, CNRS-IN2P3, 
Case 070, Place Eug\`ene
Bataillon, 34095 - Montpellier, France.}
   \ead{snarison@yahoo.fr}

%\maketitle
%\pagestyle{myheadings}
\markright{Spectra and Decay Constants of $B_c$-like and $B^*_0$ Mesons in QCD}
\begin{abstract}
\noindent
Using the existing state of art of the QCD expressions of the two-point correlators into the Inverse Laplace sum rules (LSR) within stability criteria, we  present a first analysis of the spectra and decay constants of $B_c$-like scalar $(0^{++})$ and axial-vector $(1^{++})$ mesons and revisit the ones of the $B^*_c(1^{--})$ vector meson. Improved predictions are obtained by combining these LSR results with some mass-splittings from Heavy Quark Symmetry (HQS). We complete the analysis by revisiting the $B^*_{0}(0^{++})$ mass which might be likely identified with the $B^*_J(5732)$ experimental candidate. 
The results for the spectra collected in Table\,2 are compared with some recent lattice and potential models ones. New estimates of the decay constants are given in Table\,3.
\end{abstract}
%% keywords
%\keywords{
 
\end{frontmatter}
%%%%%%%%%%%%%%%%%%%%%%%%%%%%%
\section{Introduction}
%%%%%%%%%%%%%%%%%%%%%%%%%%%%%
%\nin
-- QCD spectral sum rules (QSSR)\,\cite{SVZa,SVZb}\,\footnote{For some introductory books and reviews,see e.g.\,\cite{ZAKA,SNB1,SNB2,SNB3,SNB4,SNB5,SNREV15,IOFFEb,RRY,DERAF,BERTa,YNDB,PASC,DOSCH}} of the inverse Laplace-type (LSR)\,\cite{BELLa,BELLb,BECCHI,SNR} 
have been used successfully to study the masses and decay constants of different hadrons.  

-- More recently in\,\cite{SNbc20,SN19}, the $B_c$-mass has been used together with constraints from the $J/\psi$ and $\Upsilon$ sum rules\,\cite{SNmom18,SNparam,SNH12,SNH11,SNH10} to extract simultaneously and accurately the running charm and bottom quark masses with the results quoted in Table\,1 which we shall use hereafter for a consistency. 

-- In this paper,  using a similar LSR approach within the same stability criteria as in Refs.\,\cite{SNbc20,SN19}, we extend the analysis done for the $B_c(0^{--})$ meson in Refs.\,\cite{SNbc20,SN19}, to study  (for the first time) the masses and decay constants of the $B_c$-like scalar $0^{++}$ and axial-vector $1^{++}$ mesons.

-- We also revisit the mass and decay constant of the vector meson $B^*_c(1^{--})$ obtained earlier using $q^2=0$ moments and the $b$-quark pole mass to NLO in  Ref.\,\cite{SNBc,SNB1} and the recent estimates of the $B^*_c(1^{--})$ decay constant and $B^*_{0}(0^{++})$ mass using LSR in Ref.\,\cite{SNp15,SNbc15}.

-- We shall complement and improve the obtained LSR results for the masses by using some mass-splittings relations obtained from the flavour and spin independence properties based on Heavy Quark Symmetry (HQS)\,\cite{ISGUR,NEU}. These results will be compared with some recent lattice\,\cite{LATT} and potential models\,\cite{QUIGG,BAGAN} estimates.
%%%%%%%%%%%%%%%%%%%%%%%%%%%%%
\section{The QCD Inverse Laplace sum rules (LSR)}
%%%%%%%%%%%%%%%%%%%%%%%%%%%%%
\vspace*{-0.25cm}\subsection*{\b The QCD interpolating currents}
%%%%%%%%%%%%%%%%%%%%%%%%%%%%%
We shall be concerned with the following QCD interpolating current:
\bea
&\la 0|J_S(x)\vert P\ra= f_S M_S^2~:~J_S(x)\equiv (m_b-m_c)\bar cb~,\nnb\\
&\la 0|J_{H}^\mu(x)|{H}\ra=f_{H} M_{H}\epsilon^\mu~:J_{H}^\mu(x)\equiv \bar c\gamma_\mu(\gamma_5) b~,
\label{eq:fp}
\label{eq:current}
\eea
where: 
 $J_S(x) $  is the local heavy-light scalar current; $J _{H}^\mu(x)$  [$H\equiv V(A)$] is the (axial)-vector currents; $\epsilon^\mu$ is the (axial) vector polarization; $m_{c,b}$ are renormalized masses of the QCD Lagrangian; $f_S,f_H$ are the decay constants related to  the leptonic width $\Gamma [S,H\to l^+\nu_l]$ and normalised as $f_\pi=132$ MeV.
% \end{document}
%%%%%%%%%%%%%%%%%%%%%%%%%%%%%
\vspace*{-0.25cm}\subsection*{\b Form of the sum rules}
%%%%%%%%%%%%%%%%%%%%%%%%%%%%% 
We shall work with the  Finite Energy version of the QCD Inverse Laplace sum rules (LSR)  :
\beq
\hspace*{-0.6cm} {\cal L}^c_n(\tau,\mu)=\int_{(m_c+m_b)^2}^{t_c}\hspace*{-1cm}dt~t^n~e^{-t\tau}\frac{1}{\pi} \mbox{Im}~[\psi_S;\Pi_H](t,\mu)~,
\label{eq:lsr}
\eeq
and their ratios :
\beq
 {\cal R}^c_n(\tau)=\frac{{\cal L}^c_{n+1}} {{\cal L}^c_n},
\eeq
 where $\tau$ is the LSR variable, $n=0,1$ is the degree of moments, $t_c$ is  the threshold of the ``QCD continuum" which parametrizes, from the discontinuity of the Feynman diagrams, the spectral function  ${\rm Im}[\psi_S,\Pi^{(1)}_H(t,m_Q^2,\mu^2)]$   where  $\psi_S(t,m_Q^2,\mu^2)$ is the  scalar and $\Pi^{(1)}_H(t,m_Q^2,\mu^2)]$ the (axial) vector correlators defined as :
 \bea
\hspace*{-0.6cm} \psi_{S}(q^2)\hspace*{-0.cm}&=&\hspace*{-0.cm}i \hspace*{-0.15cm}\int \hspace*{-0.15cm}d^4x ~e^{-iqx}\la 0\vert {\cal T} J_{S}(x)\ga J_{S}(0)\dr^\dagger \vert 0\ra~,\nnb\\
\hspace*{-0.6cm} \Pi^{\mu\nu}_H(q^2)\hspace*{-0.cm}&=&\hspace*{-0.cm}i \hspace*{-0.15cm}\int \hspace*{-0.15cm}d^4x ~e^{-iqx}\la 0\vert {\cal T} J^\mu_H(x)\ga J^\nu_H(0)\dr^\dagger \vert 0\ra\nnb\\
\hspace*{-0.6cm}&=&\hspace*{-0.cm} -\ga g^{\mu\nu}q^2-q^\mu q^\nu\dr \Pi_H^{(1)}(q^2)+ {q^\mu q^\nu}\Pi_H^{(0)}(q^2),
 \label{eq:2-pseudo}
 \eea
 with $\Pi_H^{(1,0)}$ corresponds to the spin 1, 0 meson contributions. 
   %%%%%%%%%%%%%%%%%%%%%%%%%%%%%
\section{The QCD two-point function within the SVZ-expansion}
 %%%%%%%%%%%%%%%%%%%%%%%%%%%%%
-- Using the SVZ\,\cite{SVZa} Operator Product Expansion (OPE), the Inverse Laplace tranform of the two-point correlator can be written in the form:
\bea
\hspace*{-0.5cm}{\cal L}^c_n(\tau,\mu)\hspace*{-0.cm}&=&\hspace*{-0.cm}\int_{(m_c+m_b)^2}^{t_c}\hspace*{-0.7cm} dt e^{-t\tau}
%{dt\over t-q^2}
\hspace*{-0.0cm}\frac{t^n}{\pi} \mbox{Im}~[\psi_S;\Pi^{(1)}_H](t,\mu)\vert_{PT} +\nnb\\
\hspace*{-0.3cm}&+&\hspace*{-0.cm}\la \alpha_s G^2\ra C^{G^2}_{S,H}(\tau,\mu)+ \overline{m}_b\la \bar cc\ra C^\psi_{S,H}(\tau,\mu) %+\la g^3G^3\ra C_{G^3}(q^2,\mu)
+\cdots
\eea

--  $\mbox{Im}~[\psi_S;\Pi^{(1)}_H](t,\mu)\vert_{PT} $ is the perturbative part of the spectral function. $C^{G^2}_{S,H}$ and $C^{\psi}_{S,H}$ are (perturbatively) calculable Wilson coefficients. $ \la \alpha_s G^2\ra $ and $\la\bar cc\ra$ are the non-pertubative gluon and quark condensates where $G^2\equiv G^a_{\mu\nu}G^{\mu\nu}_a$.
 
 -- We have not retained higher dimension condensates as these contributions are negligible in the working regions. The  $\la\bar bb\ra$ condensate contribution where the $b$ is considered as a heavy quark here is already included in $C_{G^2}$  as explicitly shown in Ref.\,\cite{BAGAN} through the relation\,\cite{SVZa,GENERALIS1,BAGAN1}:
\beq
\la \bar bb\ra=-{1\over 12\pi m_b}\la\alpha_s G^2\ra+{\cal O}\ga 1/m_b^3\dr +\cdots~.
 \eeq
 
 -- The charm quark is considered as a light quark where an expansion in $m_c^2/Q^2$ and $m_c/m_b$ has been done for the non-perturbative contributions. The corresponding condensate is estimated from the analogue previous relation with the gluon condensate. 
 
--  In the two next sections, we shall collect the QCD expressions of the two-point correlators known to NLO and N2LO in the literature from which we shall derive the
 expressions of the sum rules.
 %%%%%%%%%%%%%%%%%%%%%%%%%%%%%
\section{The $q^2=0$ behaviour of the two-point function}
%%%%%%%%%%%%%%%%%%%%%%%%%%%%% 
 %%%%%%%%%%%%%%%%%%%%%%%%%%%%%%%%%%%%%%%%%%%
 To NLO, the perturbative part of $\psi_{S}(0)$ reads\,\cite{SNB1,SNB2,BECCHI,GENERALIS}\,\footnote{Analogous relation between the correlators of the pseuscalar and axial currents has been already discussed in Ref.\,\cite{BECCHI,SNbc20}.}
 \beq
 \psi_{S}(0)\vert_{PT}={3\over 4\pi^2}(m_b-m_c)\ga m_b^3Z_b+m_c^3Z_c\dr~,
 \eeq
 with : 
 \beq
 Z_i=\ga 1- \log{m_i^2\over \mu^2}\dr\ga1+{10\over 3}a_s\dr+{2}a_s \log^2{m_i^2\over \mu^2}~,
 \eeq
 where $i\equiv c,b$;  $\mu$ is the QCD subtraction constant and $a_s\equiv \alpha_s/\pi$ is the QCD coupling. This PT contribution which is present here has to be added to the well-known non-perturbative contribution: 
 \beq
  \psi_{S}(0)\vert_{NP}=-(m_b-m_c)\la \bar bb-\bar cc\ra~,
  \eeq
  for absorbing $\log^n(-m_i^2/q^2)$ mass singularities appearing during the  evaluation of the PT two-point function, a technical point not often carefully discussed in some papers.
Working with $\psi_{S}(q^2)$ and $\Pi_H^{(0)}$ defined previously is safe as $\psi_{S}(0),~\Pi_H^{(1)}(0)$, which disappear after successive derivatives, do not affect the sum rule. This is not the case of the  longitudinal part of the vector two-point function $\Pi^{(0)}_{V}(q^2)$ built from the vector current
which is
related to $\psi_{S}(q^2)$ through the Ward identity\,\cite{SNB1,SNB2,SNB5,BECCHI}:
 \beq
 \Pi^{(0)}_{V}(q^2)={1\over q^2}\Big{[} \psi_{S}(q^2)-\psi_{S}(0)\Big{]}~,
 \eeq
 and  which is also often (uncorrectly) used  in literature.  
 %\end{document}

 %%%%%%%%%%%%%%%%%%%%%%%%%%%%%
\section{The Two-Point Function  at large $q^2$}
%%%%%%%%%%%%%%%%%%%%%%%%%%%%% 
We have given in details the QCD expression of the pseudoscalar spectral function in Ref.\,\cite{SNbc20}.
Some (not lengthy) expressions of the other spectral functions are given below. 
%%%%%%%%%%%%%%%%%%%%%%%%%
\vspace*{-0.25cm}\subsection*{\b  Perturbative contributions}
%%%%%%%%%%%%%%%%%%%%%%%%%
-- The complete expressions  of the PT spectral function in terms of the on-shell quark masses has been obtained to LO in\,\cite{FNR}:
\bea
&&\hspace*{-0.cm}{\rm Im}\psi_{5(S)}(t)\hspace*{-0.cm}=\hspace*{-0.cm}{3\over8\pi}(m_c\pm m_b)^2t\ga 1-{(m_b\mp m_c)^2\over t}\dr \lambda^{1/2},\nnb\\
&&\hspace*{-0.cm}{\rm Im}\Pi^{(1)}_{A(V)}(t)\hspace*{-0.cm}=\hspace*{-0.cm}{3\over12\pi}\Big{[} 1-{m_b^2+m_c^2\pm 6m_cm_b\over 2t}-\nnb\\
&&\hspace*{4cm}{(m_c^2-m_b^2)^2\over 2t^2}\Big{]} \lambda^{1/2},
\label{eq:2point}
\eea
with the phase space factor :  
\beq
\lambda^{1/2}=\ga 1-{(m_b+m_c)^2\over t}\dr^{1/2}\ga 1-{(m_b-m_c)^2\over t}\dr^{1/2}.
\eeq

-- The lengthy expressions at NLO are given in Refs \,\cite{GENERALIS,RRY,BAGAN}. The ones for the states of opposite parities can be obtained by a careful change of the sign of one of the quark mass (chirality transformation due to the (non)-presence of the $\gamma_5$ Dirac matrix). 

-- We shall use the N2LO contributions obtained in the limit where one of the quark mass is zero \cite{CHETa,CHETb}  which we expect to be a good approximation as the N2LO correction is relatively small. This expression is available as  a Mathematica program Rvs.m. 

-- We estimate the error due to the truncation of the PT series from the N3LO
contribution using a geometric growth of the PT series which is expected to mimic the phenomenological $1/q^2$ dimension-two contribution\,\cite{SZ} parametrizing the uncalculated large order terms of PT series \cite{CNZa,CNZb}\,\footnote{For reviews, see e.g.\,\cite{ZAKa,ZAKb}.}.  
%%%%%%%%%%%%%%%%%%%%%%%%%%%%% 
\vspace*{-0.25cm}\subsection*{\b  Non-perturbative contributions}
%%%%%%%%%%%%%%%%%%%%%%%%%
-- The complete non-perturbative contributions due to  the gluon condensate have been obtained by\,\cite{BAGAN,RRY} at LO.  These expressions are also lengthy and will not be reported here. However, as these contributions are relatively small in the analysis, it is a good approximation to work with the approximate expressions where linear and quadratic corrections in term of $m_c$ are retained. 

 Moreover, one should be careful in using the expressions given by \,\cite{GENERALIS,NSV2Z,JAMIN1,PIVOV} for the (axial)vector 
currents as the decomposition of the correlator used there  is slightly different of the one in Eq.\,\ref{eq:2point} ($H\equiv V,A$):
\beq
\Pi_{H}^{\mu\nu}(q^2)= -\ga g^{\mu\nu}q^2-{q^\mu q^\nu}\dr \Pi_{H}^T(q^2)+ g^{\mu\nu}q^2\Pi_{H}^{(0)}(q^2)~,
\eeq

The relevant component associated to the spin 1 meson used in\,\cite{SNp15} which we shall use in the following, is the combination :
\beq
 \Pi^{(1)}_{H}(q^2)\equiv \Pi^{T}_{H}(q^2) -  \Pi^{(0)}_{H}(q^2)~.
 \eeq 
 
 To avoid singularities at $q^2=0$ (see e.g.\,\cite{GENERALIS,JAMIN1}) and some (non) perturbative effects  due to $\Pi^{(1,0)} (0)$, we shall work with the Inverse Laplace transform of the rescaled function :
 \beq
  \tilde\Pi^{(1)}_{H}(q^2)\equiv  q^2\Pi^{(1)}_{H}(q^2)~.
 \eeq
 --  The $\la \bar cc\ra$ quark condensate contribution to $\psi_S(q^2)$ is given to LO by\,\cite{GENERALIS,NSV2Z,ELETSKY} and to NLO by\,\cite{JAMIN,PIVOV}:
\bea 
\hspace*{-0.cm}C^\psi_{S}\hspace*{-0.cm}&=&\hspace*{-0.cm}(m_b- m_c)^2\Bigg{[}\Big{[} 1 + (1 + z){m_c\over2m_b} + z{m_c^2\tau\over 2}\Big{]} {\rm e}^{-z}\nnb\\
&&\hspace*{4.5cm}-{a_s\over 2}C^\psi_{S1}\Bigg{]}~:\nnb\\
\hspace*{-0.cm}C^\psi_{S1}\hspace*{-0.cm}&=&\hspace*{-0.cm}\Gamma(0,z)-\Bigg{[} 1+2(1-z)\ga l_{\mu b}+{2\over 3}\dr\Bigg{]}{\rm e}^{-z}~,
\eea
where : $z\equiv m_b^2\tau$, $ l_{\mu b}\equiv \log\ga \mu/m_b\dr$ and $\Gamma(n,z)$ is the {\it n}-th incomplete $\Gamma$-function.

-- The $\la \bar cc\ra$ quark condensate contribution to $\tilde\Pi^{(1)}_V(q^2)$ is derived from the expression given by\,\cite{GENERALIS,PIVOV}.
It reads :
\bea 
\hspace*{-0.cm}C^\psi_{V}\hspace*{-0.cm}&=&\hspace*{-0.cm}
-e^{-z}\, \Bigg{[} 1 -{m_c m_b}{\tau\over 2}+{2\over 3}a_s C^\psi_{V1}\Bigg{]}:
\nnb\\
\hspace*{-0.cm}C^\psi_{V1}\hspace*{-0.cm}&=&\hspace*{-0.cm} 1-6z l_{\mu b}-4z+\Gamma(-1,z)\,z\,{\rm e}^{-z}~.
\eea

-- The gluon condensate contribution reads to LO :
\bea
\hspace*{-0.cm}C^{G^2}_S\hspace*{-0.cm}&=&\hspace*{-0.cm}{(m_b - m_c)^2\over 12\pi}
\Bigg{[} {\rm e}^{-z}-\ga{m_c\over m_b}\dr C^{G^2}_{S1}+\ga{m_c\over m_b}\dr^2C^{G^2}_{S2}\Bigg{]} ~: \nnb\\
\hspace*{-0.cm}C^{G^2}_{S1}\hspace*{-0.cm}&=&\hspace*{-0.cm}\Bigg{[} 1+2z-3z^2\ga1+ l_{\mu b}\dr\Bigg{]} e^{-z}
-6f_3(z),\nnb\\
%-3z^2\ga \rm{log}{\tau\mu^2}-\psi(3)\dr ,\nnb\\
\hspace*{-0.cm}C^{G^2}_{S2}\hspace*{-0.cm}&=&\hspace*{-0.cm}\Bigg{[} 1+z+z^2-{z^3}\ga {7\over 6}+ l_{\mu b}\dr\Bigg{]} e^{-z} -6f_4(z)~,\nnb\\
%- z^3\ga\log{\tau\mu^2}-\psi(4)\dr,\nnb\\
\hspace*{-0.cm}C^{G^2}_V\hspace*{-0.cm}&=&\hspace*{-0.cm}-{1\over 12\pi}
\Bigg{[} {\rm e}^{-z}-\ga{m_c\over m_b}\dr C^{G^2}_{V1}+\ga{m_c\over m_b}\dr^2C^{G^2}_{V2}\Bigg{]}: \nnb\\
\hspace*{-0.cm}C^{G^2}_{V1}\hspace*{-0.cm}&=&\hspace*{-0.cm}C^{G^2}_{S1} ,\nnb\\
\hspace*{-0.cm}C^{G^2}_{V2}\hspace*{-0.cm}&=&\hspace*{-0.cm}\Bigg{[} 1+{11\over 3}z-{22\over3}z^2+{7\over 6}{z^3} -(6-7z)z^2 l_{\mu b}\Bigg{]} e^{-z}\nnb\\
&&-g_4(z)~,\nnb\\
%&&-{\cal L}[(1-2x)f_4(x)],
\eea
where the $m_c=0$ result comes from\,\cite{NSV2Z,ELETSKY,GENERALIS,JAMIN,PIVOV}. 

-- The $m_c$-corrections have been derived from the expression of the two-point correlators given by\,\cite{GENERALIS}. The functions  $f_n(z)$ and $g_n(x)$ are respectively the Inverse Laplace transform of the functions $Y^n(x)\log(x)$ and  $(1+2/x)Y^n(x)\log(x)$ with $ Y(x)\equiv 1/(1+1/x)$ and $x\equiv m_b^2/Q^2$.

-- The expressions of the correlators associated to the pseudoscalar and axial-vector currents can be deduced from the former by the chiral transformation : 
\beq
m_c\to -m_c. 
\eeq
%%%%%%%%%%%%%%%%%%%%%%%%%%%%%
\vspace*{-0.25cm}\subsection*{\b  From the On-shell to the $\overline{MS}$-scheme}
%%%%%%%%%%%%%%%%%%%%%%%$%%%%%% 
We transform the pole masses $m_Q$ to the running masses $\overline m_Q(\mu)$ using the known relation  in the
%the $\overline{MS}$-scheme :
%-- Using the known relation between the running $\bar{m}_Q(\mu)$ and on-shell mass
%$M_Q$ in the 
$\overline{MS}$-scheme to order $\alpha_s^2$ \cite{TAR,COQUEa,COQUEb,SNPOLEa,SNPOLEb,BROAD2a,BROAD2b,CHET2a,CHET2b}:
\bea
m_Q &=& \overline{m}_Q(\mu)\Big{[}
1+{4\over 3} a_s+ (16.2163 -1.0414 n_l)a_s^2\nnb\\
&&+\ln{\mu^2\over m_Q^2} \ga a_s+(8.8472 -0.3611 n_l) a_s^2\dr\nnb\\
&&+\ln^2{\mu^2\over m_Q^2} \ga 1.7917 -0.0833 n_l\dr a_s^2...\Big{]},
\label{eq:pole}
\eea
for $n_l=3: u,d,s$ light flavours. In the following, we shall use $n_f=5$ total number of flavours for the numerical value of $\alpha_s$. 
%%%%%%%%%%%%%%%%%%%%%%%%%%%%%%%%%%%
\section{QCD input parameters}
%%%%%%%%%%%%%%%%%%%%%%%%%%%%%%%%%%%
%\nin
The QCD parameters which shall appear in the following analysis will be the QCD coupling $\alpha_s$, the charm and bottom running quark masses $m_{c,b}$ and
 the gluon condensate $ \la\alpha_sG^2\ra$.  Their values are given in Table\,1. 
% \end{document}
 %%%%%%%%%%%%%%%%%%%%%%%%%%%%%%%%%%%%%%%%%%

\begin{table}[H]
%\label{tab:param}
{\scriptsize
 \caption{QCD input parameters from recent QSSR analysis based on stability criteria.
 $\overline{m}_{c,b}(\overline {m}_{c,b})$ are the running $c,b$ quark masses evaluated at $\overline{m}_{c,b}$. }  }
%\tbl{
%}
\setlength{\tabcolsep}{0.2pc}
    {\small
 \begin{tabular}{llll}
% {\begin{tabular}{@{}llll@{}} \toprule
&\\
\hline
\hline
%\\
Parameters&Values&Sources& Ref.    \\
%\\
\hline
$\alpha_s(M_Z)$& $0.1181(16)(3)$&$M_{\chi_{0c,b}-M_{\eta_{c,b}}}$&LSR \, \cite{SNparam}\\
$\overline{m}_c(\overline {m}_c)$&$1286(16)$ MeV &$B_c\oplus {J/\psi}$&Mom.\,\cite{SNbc20,SNmom18}\\
$\overline{m}_b(\overline {m}_b)$&$4202(8)$ MeV&$B_c\oplus{\Upsilon}$&Mom.\,\cite{SNbc20,SNmom18}\\
%$\hat \mu_q$&$(253\pm 6)$ MeV&\cite{SNB1,SNmassa,SNmassb,SNmass98a,SNmass98b,SNLIGHT}\\
%$\la \bar dd\ra(2) $&$-(275.7\pm 6.6)^3$ MeV$^3$&\cite{SNB1,SNmass}\\
%$M_0^2$&$(0.8 \pm 0.2)$ GeV$^2$&\cite{JAMI2a,JAMI2b,JAMI2c,HEIDa,HEIDb,HEIDc,SNhl}\\
$\la\alpha_s G^2\ra$& $(6.35\pm 0.35)\times 10^{-2}$ GeV$^4$&Hadrons&Average\,\cite{SNparam}\\
%$\la g^3  G^3\ra$& $(8.2\pm 2.0)$ GeV$^2\times\la\alpha_s G^2\ra$&$J/\psi$  family&Mom. \cite{SNH10a,SNH10b}\\
%&&&Ratios of LSR \cite{SNH10c}\\
%$\rho \alpha_s\la \bar qq\ra^2$&$(5.8\pm 1.8)\times 10^{-4}$ GeV$^6$&\cite{SNTAU,LNT,JAMI2a,JAMI2b,JAMI2c}\\
%$\hat m_s$&$(0.114\pm0.006)$ GeV &\cite{SNB1,SNTAU9,SNmassa,SNmassb,SNmass98a,SNmass98b,SNLIGHT}\\
%$\kappa\equiv \la \bar ss\ra/\la\bar dd\ra$& $(0.74^{+0.34}_{- 0.12})$&\cite{HBARYONa,HBARYONb,SNB1}\\
\hline\hline
\end{tabular}
}
\label{tab:param}
%\caption{%\scriptsize   
\end{table}

%%%%%%%%%%%%%%%%%%%%%%%%%%%%%%
\vspace*{-0.25cm}\subsection*{\b  QCD coupling $\alpha_s$}
%%%%%%%%%%%%%%%%%%%%%%%%%%%%%%%
We shall use the value of $\alpha_s$ from the $M_{\chi_{0c}}-M_{\eta_{c}}$ mass-splitting sum rule\,\cite{SNparam}: 
 \bea&&\hspace*{-1cm} 
 \alpha_s(2.85)=0.262(9) \lrar\alpha_s(M_\tau)=0.318(15)\nnb\\
& \lrar&\alpha_s(M_Z)=0.1183(19)(3)
 %\label{eq:muc}
\eea
which is more precise than the one from  $M_{\chi_{0b}}-M_{\eta_{b}}$\,\cite{SNparam}\,: 
\bea &&\hspace*{-1cm} 
 \alpha_s(9.50)=0.180(8) \lrar\alpha_s(M_\tau)=0.312(27)\nnb\\
&&\lrar\alpha_s(M_Z)=0.1175(32)(3).
 \eea
 These lead to the mean value quoted in Table\,1, which is
 in agreement with the one from $\tau$-decays \,\cite{PICH,SNTAU} and with the world average\,\cite{PDG}:
 %,BETHKE,PICH,SALAM}: 
\beq
\alpha_s(M_\tau)=0.325(8)~~{\rm and}~~\alpha_s(M_Z)\vert_{\rm average}=0.1181(11),
\eeq
but with a larger error. 
%%%%%%%%%%%%%%%%%%%%%%%%%%%%%%
\vspace*{-0.25cm}\subsection*{\b  $c$ and $b$ quark masses}
%%%%%%%%%%%%%%%%%%%%%%%%%%%%%%%
For the $c$ and $b$ quarks, we shall use the recent determinations\,\cite{SNmom18,SNbc20} of  the running masses and the corresponding value of $\alpha_s$ evaluated at the scale $\mu$ obtained using the same sum rule approach from charmonium and bottomium systems. These values are quoted in Table\,1.
%%%%%%%%%%%%%%%%%%%%%%%%%%%%%%
\vspace*{-0.25cm}\subsection*{\b  Gluon  condensate $\la \alpha_s G^2\ra$}
%%%%%%%%%%%%%%%%%%%%%%%%%%%%%%%
We use the recent QSSR average from different channels\,\cite{SNparam} quoted in Table\,1 which includes the recent estimate obtained from a correlation with the values of the heavy quark masses and $\alpha_s$.  

%%%%%%%%%%%%%%%%%%%%%%%%%%%%%
\section{Parametrisation of the spectral function}
%%%%%%%%%%%%%%%%%%%%%%%%%%%%% 
-- In the present case, where no complete data on the spectral function are available, we use the duality ansatz:
 \bea
\hspace*{-0.25cm} {\rm Im} [\psi_S;\Pi_H^{(0)}]\hspace*{-0.cm}&\simeq&\hspace*{-0.cm} f_H^2 M_H^{2p} \delta(t-M_H^2)+\nnb\\
&& \Theta(t-t_c) ``{\rm  QCD continuum}",
 \eea
 for parametrizing the spectral function. $M_H$ and $f_H$ are the lowest ground state mass and coupling analogue to $f_\pi$ where $p=0$ for $H\equiv V,A$ and $p=2$ for $H\equiv S$. The ``QCD continuum"  is the imaginary part of the QCD correlator from the  threshold $t_c$. Within a such parametrization, one  obtains: 
 \beq
  {\cal R}^{c}_n\equiv {\cal R}\simeq M_H^2~,
  \label{eq:mass}
  \eeq
 indicating that the ratio of moments appears to be a useful tool for extracting the mass of the hadron ground state\,\cite{SNB1,SNB2,SNB3,SNB4,SNREV15}. 
 
 -- This simple model has been tested in different channels where complete data are available (charmonium, bottomium and $e^+e^-\to I=1$ hadrons)\,\cite{SNB1,SNB2,BERTa}. It was shown that, within the model, the sum rule  reproduces well the one using the complete data, while
the masses of the lowest ground state mesons ($J/\psi,~\Upsilon$ and $\rho$) have been predicted with a good accuracy.  In the extreme case of the Goldstone pion, the sum rule using the spectral function parametrized by this simple model\,\cite{SNB1,SNB2} and the more complete one by ChPT\,\cite{BIJNENS} lead to  similar values of the sum of light quark masses $(m_u+m_d)$ indicating the efficiency of this simple parametrization. 

-- An eventual violation of the quark-hadron duality (DV)\,\cite{SHIF,PERIS} has been frequently tested  in the accurate determination of $\alpha_s(\tau)$ from hadronic $\tau$-decay data\,\cite{PERIS,SNTAU,PICH}, where its quantitative effect in the spectral function was found to be less than 1\%. Typically, the DV behaves as: 
\beq
\Delta{\rm Im} [\psi_S;\tilde\Pi^{(1)}_H](t)\sim t~{\rm e}^{-\kappa t} {\rm sin} (\alpha+\eta t) \theta (t-t_c)~,
\eeq 
where $\kappa,\alpha,\eta$ are model-dependent  fitted parameters but not based from first principles. Within this model, where the contribution is doubly exponential suppressed in the Laplace sum rule analysis, we expect that in the stability regions where the QCD continuum contribution to the sum rule is minimal and where the optimal results in this paper will be extracted, such duality violations can be safely neglected. 

-- Therefore, we (a priori) expect that one can extract with a good accuracy the  masses and  decay constants of the $B_c$-like mesons within the approach. An eventual improvement of the results can be done after a more complete measurement of the $B_c$-like spectral function which is not an easy experimental task. 
% though the recent discovery by CMS\,\cite{CMS} of the $B_c(2S)$ state at 6872(1.5)\,MeV is a good starting point in  this direction. 
 
-- In the following, in order to minimize the effects of unkown higher radial excitations smeared by the QCD continuum and some eventual quark-duality violations, we shall work with the lowest ratio of moments $ {\cal R}^{c}_0$ for extracting the meson masses and with the lowest moment $ {\cal L}^c_0 $ for estimating the decay constant $f_{H}$. Moment with negative $n$ will not be considered due to their sensitivity on the non-perturbative contributions such as $\psi_S(0)$. 
  
 %%%%%%%%%%%%%%%%%%%%%%%%%%%%%
\section{Optimization Criteria}
%%%%%%%%%%%%%%%%%%%%%%%%%%%%% 
-- For extracting the optimal results from the analysis, we have used in previous works the optimization criteria (minimum sensitivity) of the observables versus the variation  of the external variables namely the $\tau$-sum rule parameter, the QCD continuum threshold $t_c$ and the subtraction point $\mu$. 

-- Results based on these criteria have lead to successful predictions in the current literature\,\cite{SNB1,SNB2}. $\tau$-stability has been introduced and tested by Bell-Bertlmann using the toy model of harmonic oscillator\,\cite{BERTa} and applied successfully in the heavy\, \cite{BELLa,BELLb,BERTa,BERTb,BERTc,BERTd,NEUF,SHAW,SNcb3,SNHeavy,SNHeavy2,SNHQET13}  and light quarks systems\,\cite{SVZa,SVZb,SNB1,SNB2,SNB3,SNB4,SNREV15,SNL14}. 

-- It has been extended later on to the $t_c$-stability\,\cite{SNB1,SNB2,SNB3,SNB4} and  to the $\mu$-stability criteria\,\cite{SNp13,SNHQET13,SNL14,SNp15,SNparam}. 

-- Stability on the number $n$ of heavy quark moments have also been used\,\cite{SNH12,SNH11,SNH10,SNmom18}. 

-- One should notice in the previous works that these criteria have lead to more solid theoretical basis and noticeable improvement of the sum rule results. The quoted errors in the results are conservative as the range covered by $t_c$ from the beginning of $\tau$-stability to the one of $t_c$-stability is quite large. However, such large errors induces less accurate predictions compared with some other approaches (potentiel models. lattice calculations) especially for the masses of the mesons. 
This is due to the fact that, in most cases, there are no available data for heavy-light radial excitations which can used to restrict the range of $t_c$-values.  

-- However, one should note that the value of $t_c$ used in the ``QCD continuum" model does not necessarily co\"\i ncide with the 1st radial excitation mass as the "QCD continuum" is expected to smear all higher states contributions to the spectral function. This feature has been explicitly verified by\,\cite{LAUNERb} in the $\rho$-meson channel. In the case of the $B_c$ meson, we have seen\,\cite{SNbc20} that the optimal result has been obtained for $\sqrt{t_c}\simeq (7.8-8.4)$ GeV which is about 1 GeV above  the recent $B_c(2S)$-mass found at 6872(1.5) MeV by CMS\,\cite{CMS}. 

-- In order to slightly restrict the large range of variations of $t_c$ and to minimize the dependence on the form of the  ``QCD continuum" model, we shall require that its contribution to the spectral function does not exceed (20-25)\% of the lowest resonance one. 
%We shall see that this requirement will coinc\"\i ncide with range of $t_c$-values corresponding to 
%%%%%%%%%%%%%%%%%%%%%%%%%%%%%
 %%%%%%%%%%%%%%%%%%%%%%%%%%%%%
\section{$0^{++}$ Scalar channel}
%%%%%%%%%%%%%%%%%%%%%%%%%%%%% 
%%%%%%%%%%%%%%%%%%%%%%%%%%%%%
The analysis here and in the follwing sections is very similar to the case of pseusoscalar channel studied in details in\,\cite{SNbc20}. The results are summarized in different figures.
%%%%%%%%%%%%%%%%%%%%%%%%%%%%%%%%%%%
\vspace*{-0.25cm}\subsection*{\b  $\tau$-stability}
%%%%%%%%%%%%%%%%%%%%%%%%%%%%%%%%%%%
 %%%%%%%%%%%%%%%%%%%%%%%%%%%%%%%%%%%%%%%
\begin{figure}[hbt]
\vspace*{-0.25cm}
\begin{center}
%\centerline {\hspace*{-7.5cm} \bf a) }
\includegraphics[width=11cm]{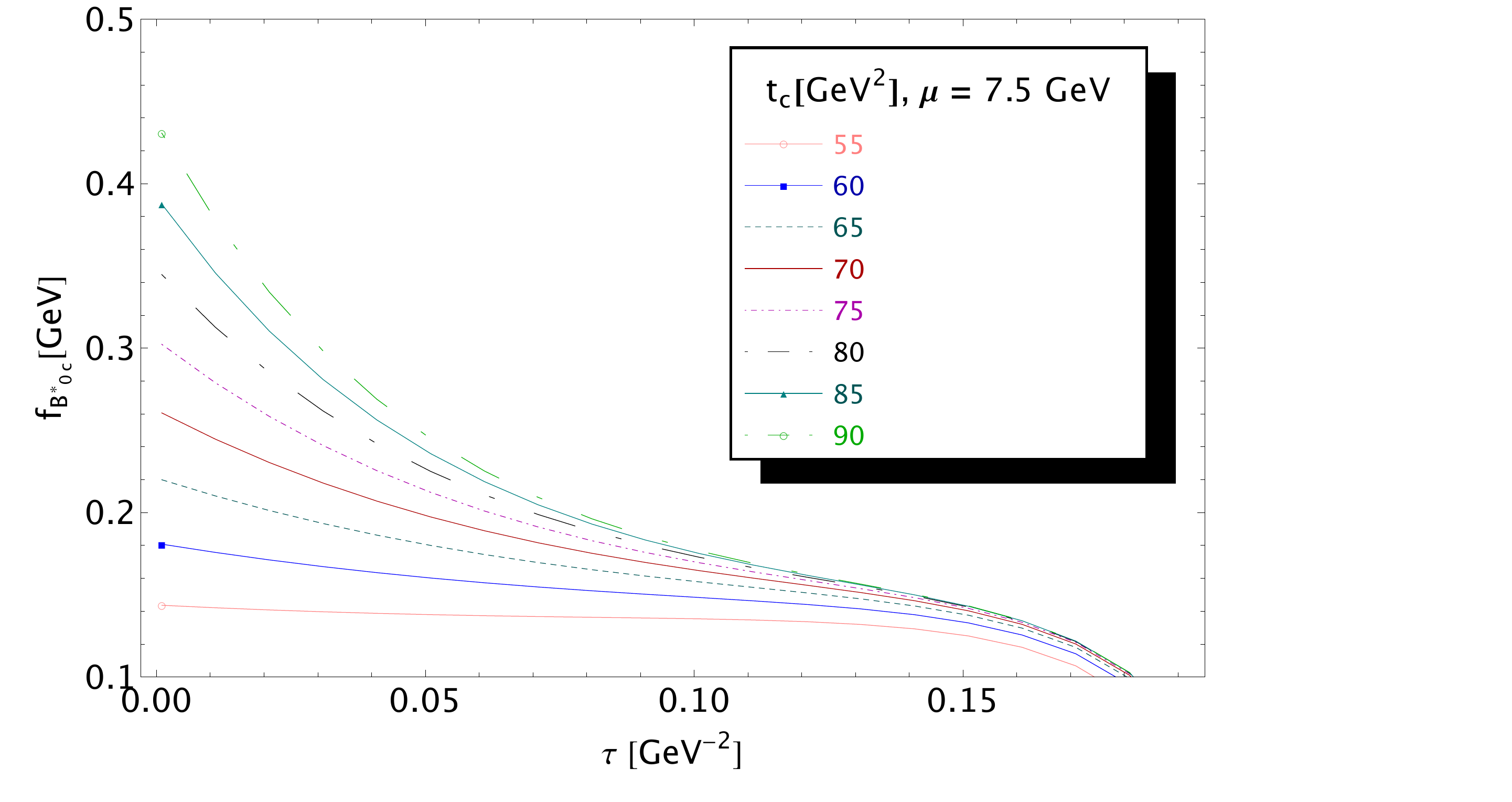}
%\vspace{0.25cm}
%\centerline {\hspace*{-7.5cm} \bf b) }
\includegraphics[width=11.2cm]{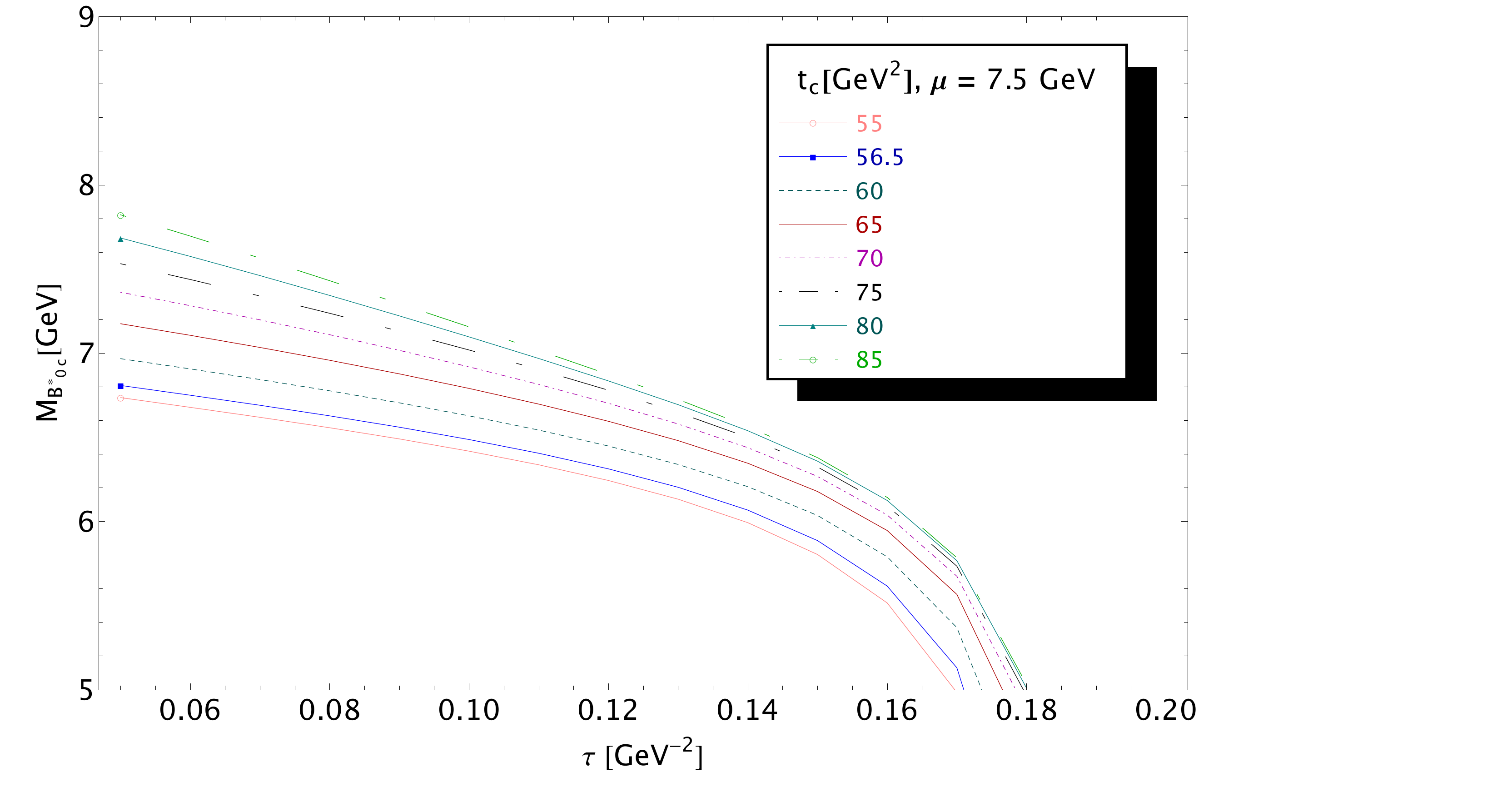}
\vspace*{-0.5cm}
\caption{\footnotesize  $f_{B^*_{0c}}$ and $M_{B^*_{0c}}$ as function of $\tau$ for different values of $t_c$, for $\mu$=7.5 GeV and for values of $\overline m_{c,b}(\overline m_{c,b})$ given in Table\,1.} 
\label{fig:bcstau}
\end{center}
%\vspace*{-0.75cm}
\end{figure} 
%%%%%%%%%%%%%%%%%%%%%%%%%%%%%%%%%%%%%%%
In a first step, fixing  the value of $\mu=7.5$ GeV which we shall justify later and which is the central value obtained in\,\cite{SNbc20,SNp15}, we show in Fig.\,\ref{fig:bcstau} the $\tau$-behaviour of $f_{B_{c^*0}}$ and $M_{B_{c^*0}}$ for different values of $t_c$ where the central values of $\overline{m}_{c,b}(\overline{m}_{c,b})$ given in Table\,1 have been used.
We see that $f_{B_{c^*0}}$ but not $M_{B_{c^*0}}$ presents inflexion points  at $\tau\simeq (0.11-0.12)$ GeV$^{-2}$ which appear for $t_c\geq$ 55 GeV$^2$. We shall use these inflexion points to fix the values of $M_{B_{c^*0}}$.
%%%%%%%%%%%%%%%%%%%%%%%%%%%%%%%%%%%%%%%
\begin{figure}[hbt]
\vspace*{-0.25cm}
\begin{center}
%\centerline {\hspace*{-7.5cm} \bf a) }
\includegraphics[width=9.5cm]{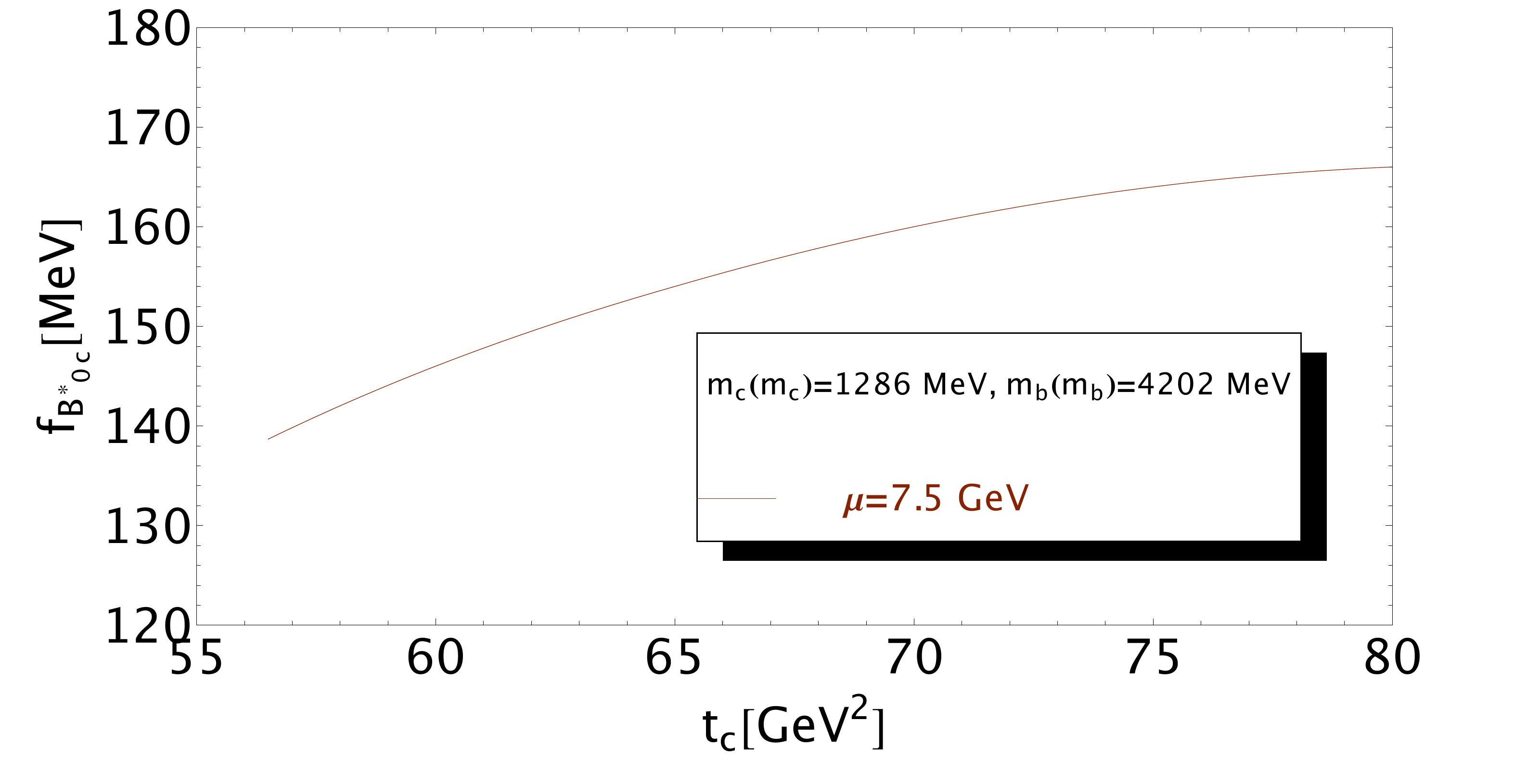}
%\centerline {\hspace*{-7.5cm}\bf b) }\\
%\vspace{0.25cm}
\includegraphics[width=9.5cm]{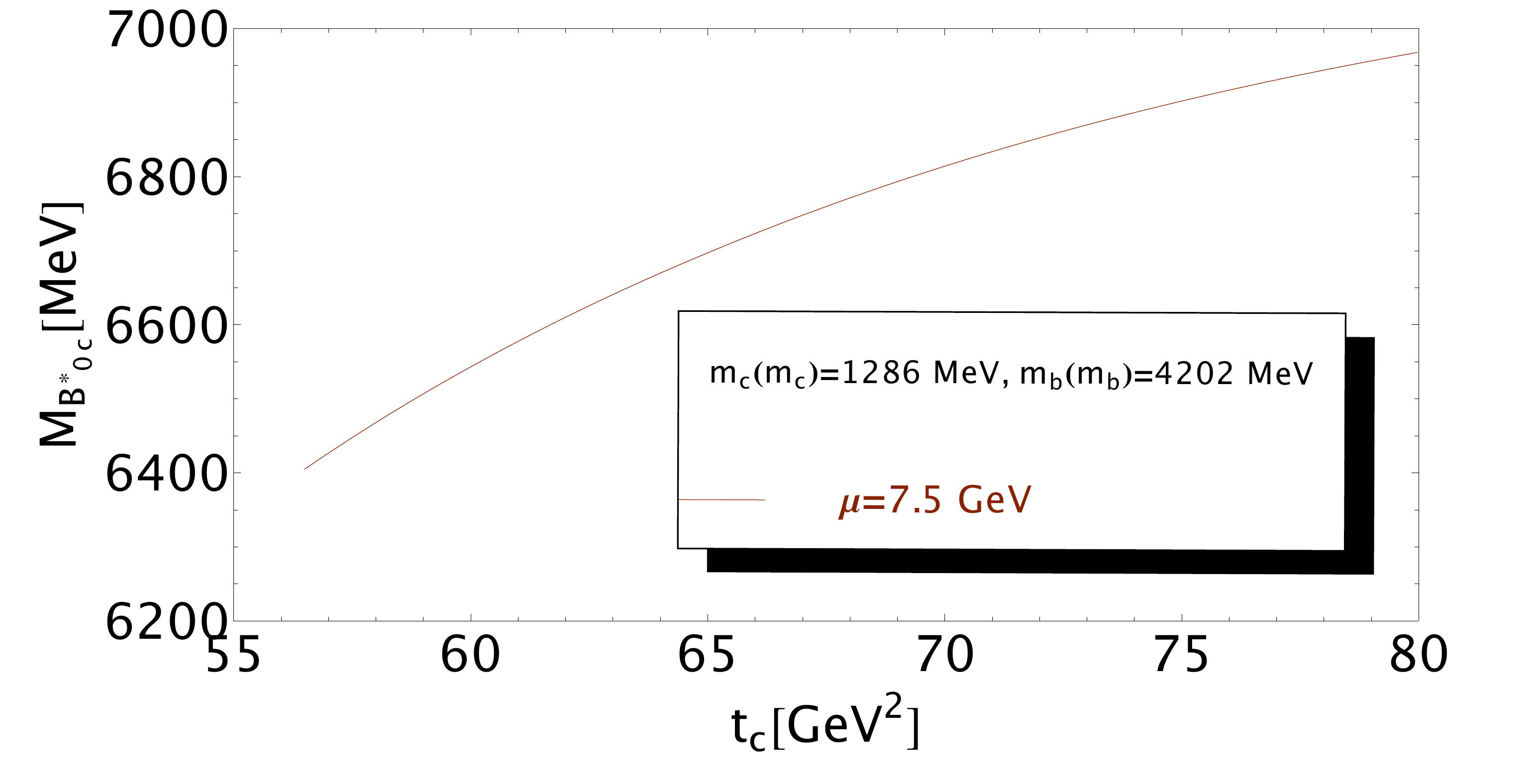}
\vspace*{-0.5cm}
\caption{\footnotesize  $f_{B_{c^*0}}$ and $M_{B_{c^*0}}$ as function of $t_c$ for $\mu$=7.5 GeV and for $\tau\simeq 0.11$ GeV$^{-2}$.} 
\label{fig:mbcs-tc}
\end{center}
%\vspace*{-0.75cm}
\end{figure} 
%%%%%%%%%%%%%%%%%%%%%%%%%%%%%%%%%%%%%%%
%%%%%%%%%%%%%%%%%%%%%%%%%%%%%%%%%%%
\vspace*{-0.25cm}\subsection*{\b  $t_c$-stability}
%%%%%%%%%%%%%%%%%%%%%%%%%%%%%%%%%%%
We study the $t_c$-behaviour of  of $f_{B_{c^*0}}$ and $M_{B_{c^*0}}$ in Fig.\,\ref{fig:mbcs-tc} where we see that $f_{B_{c^*0}}$ starts to stabilize from $t_c=70$ GeV$^2$. 
 %%%%%%%%%%%%%%%%%%%%%%%%%%%%%%%%%%%%%%%
\begin{figure}[hbt]
\vspace*{-0.25cm}
\begin{center}
\includegraphics[width=9.5cm]{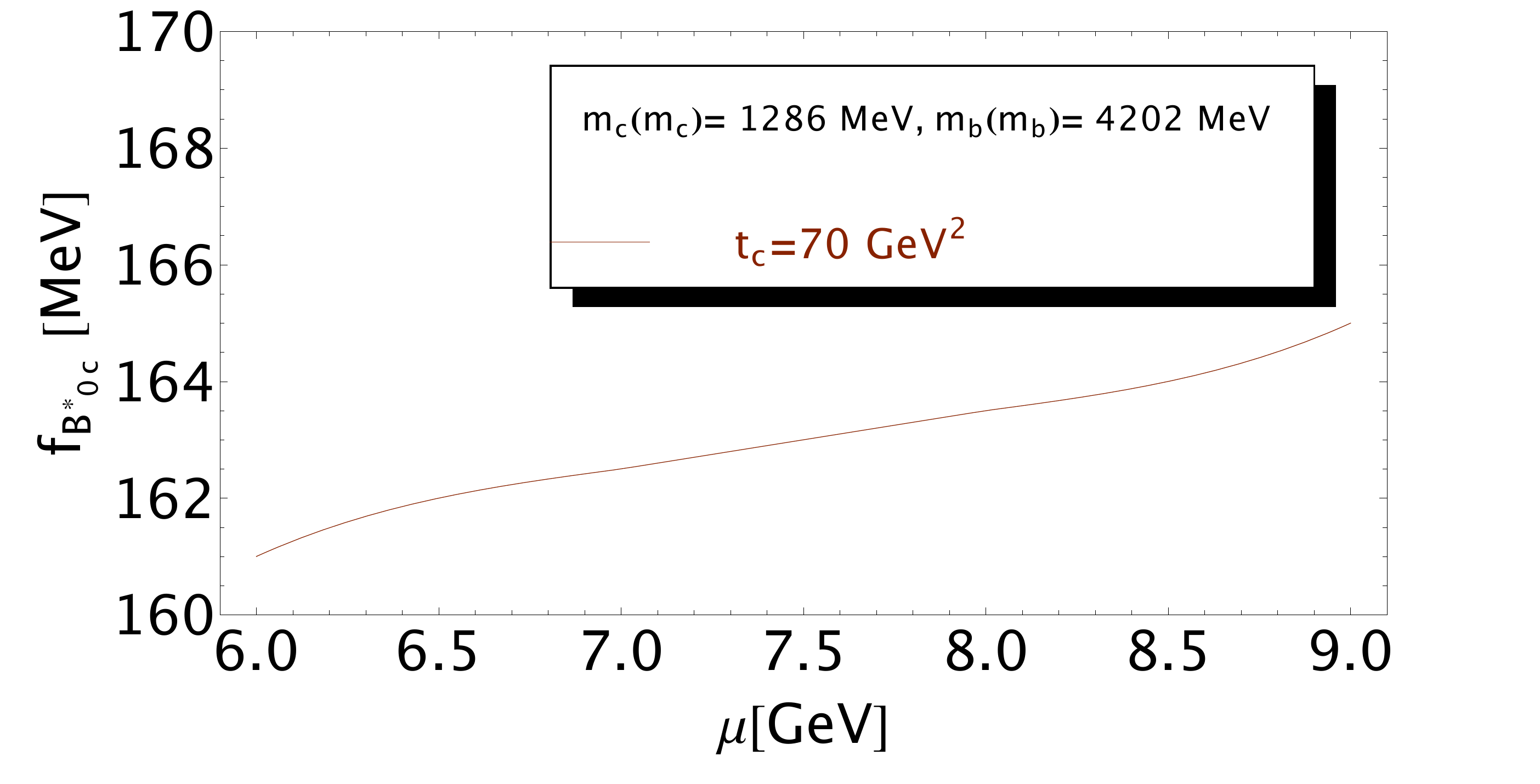}
\vspace*{-0.5cm}
\caption{\footnotesize  $f_{B_{c^*0}}$ as function of $\mu$ for $\tau\simeq 0.11$ GeV$^{-2}$ at given $t_c$.} 
\label{fig:fbcs-mu}
\end{center}
%\vspace*{-0.75cm}
\end{figure} 
%%%%%%%%%%%%%%%%%%%%%%%%%%%%%%%%%%%
\vspace*{-0.25cm}
\vspace*{-0.25cm}\subsection*{\b  $\mu$-stability}
%%%%%%%%%%%%%%%%%%%%%%%%%%%%%%%%%%%
Fixing $t_c=70$ GeV$^2$ and $\tau=0.11$ GeV$^{-2}$, we show in Fig.\,\ref{fig:fbcs-mu} the $\mu$ behaviour of $f_{B_{c^*0}}$, where we note an inflexion point at :
\beq
\mu = (7.5\pm 0.5) ~{\rm GeV}~,
\label{eq:mu}
\eeq
in agreement with the one quoted in\,\cite{SNp15,SNbc20} using different ways and/or from different channels. The $\mu$-behaviour of $M_{B_{c^*0}}$ is not shown as it is almost constant in this range of $t_c$-values. 
%%%%%%%%%%%%%%%%%%%%%%%%%%%%%%%%%%%
\vspace*{-0.25cm}\subsection*{\b  QCD continuum versus lowest resonance contribution}
\begin{figure}[hbt]
\vspace*{-0.25cm}
\begin{center}
\includegraphics[width=10cm]{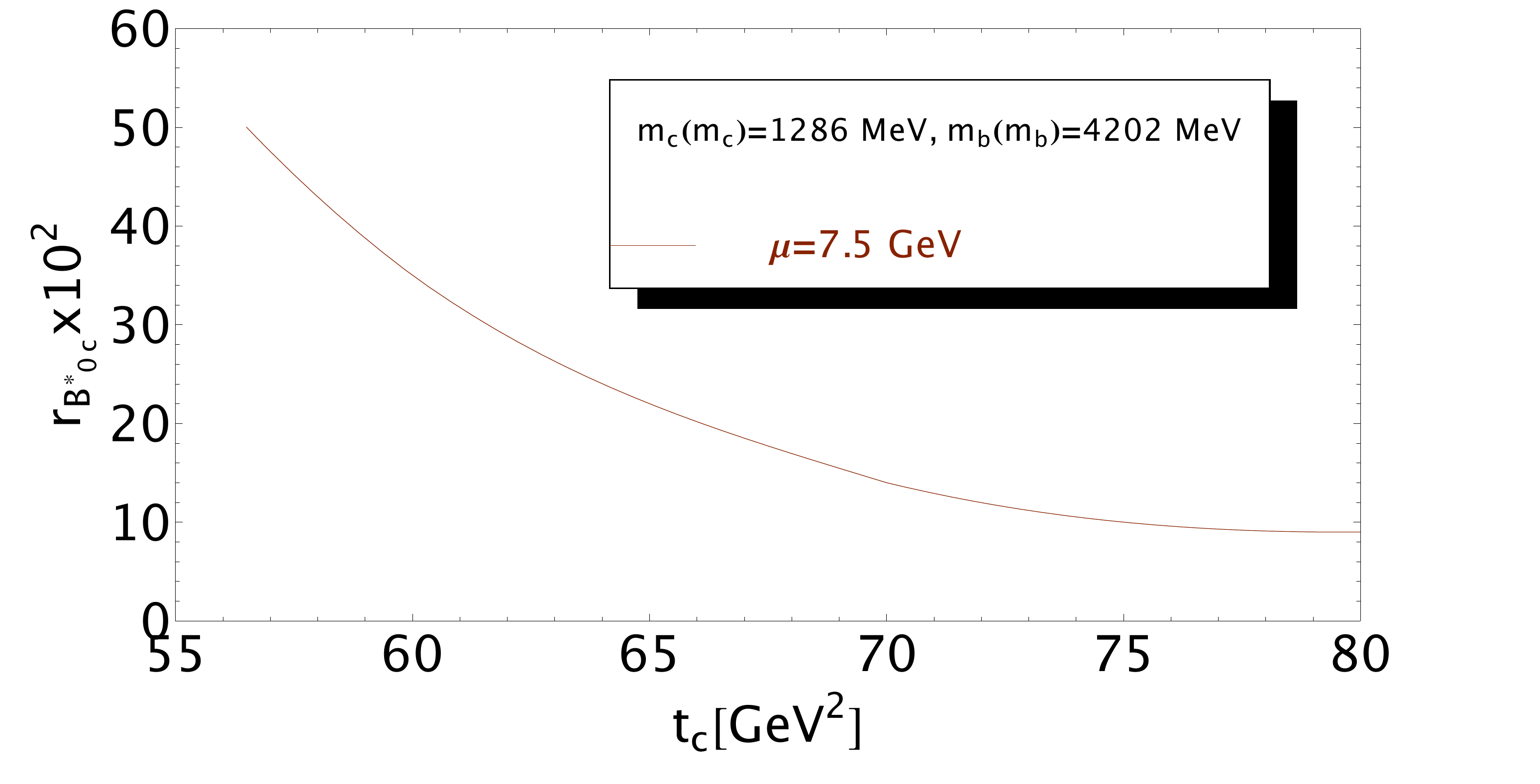}
\vspace*{-0.5cm}
\caption{\footnotesize  Ratio $r_{B_{c^*0}}$ of the continuum over the lowest ground state contribution as function of $t_c$ at the corresponding $\tau$-inflexion point for $ \mu$=7.5 GeV.} 
\label{fig:cont}
\end{center}
\end{figure}

%%%%%%%%%%%%%%%%%%%%%%%%%%%%%%%%%%%
To have more insights on the QCD continuum contribution, we show in Fig.\,\ref{fig:cont} the ratio of the continuum over the lowest ground state contribution as predicted by QCD :
\beq
r_{B_{c^*0}}\equiv{\int_{t_c}^\infty dt{\rm e}^{-t\tau}{\rm Im} \psi_S^{cont}\over\int_{(m_c+m_b)^2}^{t_c}dt{\rm e}^{-t\tau}{\rm Im} \psi_S^{B_{c^*0}}}
\eeq
%%%%%%%%%%%%%%%%%%%%%%%%%%%%%%%%%%%%%%%
%%%%%%%%%%%%%%%%%%%%%%%%%%%%%%%%%%%%%%%
The curve started from $t_c=56.5$ GeV$^2$ where the QCD continuum contribution to the spectral function is half of the resonance contribution.
One can also note from Fig.\,\ref{fig:bcstau} that the $\tau$-stability is reached from this value.
%%%%%%%%%%%%%%%%%%%%%%%%%%%%%%%%%%%
%\vspace*{-0.25cm}
\vspace*{-0.25cm}\subsection*{\b  Predictions for $M_{B_{c^*0}}$ and $f_{B_{c^*0}}$}
%%%%%%%%%%%%%%%%%%%%%%%%%%%%%%%%%%%
-- From the previous analysis and taking the large range of $t_c$ from 56.5 GeV$^2$ where the $\tau$-stability starts and where the QCD continuum is less than 50\% of the resonance one to $t_c$=75 GeV$^2$ where the $t_c$-stability is (almost) reached at which the lowest resonance dominates the spectral function (Meson Dominance Model), we obtain, for $\tau\simeq 0.11$ GeV$^{-2}$, the conservative range of predictions in units of MeV\,:
\beq
\hspace*{-0.cm}M_{B_{c^*0}}\simeq (6400-6965)~{\rm and} ~~f_{B_{c^*0}}\simeq (135-168).
\eeq

-- To improve these results, we request that the QCD continuum contribution to the spectral function is less than (20-25)\% of the resonance one. In this way, the $t_c$-values is restricted to be $(70\pm 5)$ GeV$^2$. %which is quite reasonable compared to the one $t_c=(65\pm 5)$GeV$^2$ obtained for optimizing the $B_c$-results\,\cite{SNbc20}. 
Then, we deduce the improved predictions for $\tau\simeq (0.10-0.12)$ GeV$^{-2}$ in units of MeV :
\bea
 \hspace*{-0.25cm}M_{B_{c^*0}}\hspace*{-0.1cm}&\simeq&\hspace*{-0.1cm} 6689(146)_{t_c}(112)_\tau (0)_\mu (11)_{m_{b,c}}(15)_{\alpha_s}(19)_{G^2}(67)_{syst},\nnb\\
  \hspace*{-0.25cm}f_{B_{c^*0}}\hspace*{-0.1cm}&\simeq&\hspace*{-0.1cm}155(15)_{M_{B_{c^*0}}}(4)_{t_c}(5)_\tau(0.5)_\mu(3)_{m_{b,c}}(5)_{\alpha_s}(0)_{G^2}. %(5)_{syst}\, 
  \label{eq:bcs-res} 
  \eea
  
--  We test the accuracy of the approximate expression expanded up to order $m_c^2$ by taking the example of the pseudoscalar channel where the complete expression of the non-perturbative contribution is used. We notice that the approximate result overestimates the mass prediction by about 1.01\%. We take into account this effect by adding to the prediction in Eq.\,\ref{eq:bcs-res} a systematic error of about 1\% to the mass prediction and dividing by 1.01 the estimate from the analysis.\,\footnote{Here and in the following, the quoted results for the masses take already into this systematic effect.}
   
-- Examining the analytical form of the (pseudo)scalar sum rules, one can deduce the approximate LO relation:
 \beq
f_{B_c} \simeq f_{B_{c^*0}} \ga{\overline {m}_b+\overline {m}_c\over \overline {m}_b-\overline {m}_c}\dr\ga{M_{B_{c^*0}}\over M_{B_c}}\dr^2\hspace*{-0.1cm}\times\,\rho
%e^{\ga M_{B_{c^*0}}^2-M_{B_{c^*0}}^2\dr{\tau\over 2}}
\approx(348\sim 468)~{\rm MeV},
\eeq
where $\rho\equiv {\rm Exp}[{{(M_{B_{c}}^2-M_{B_{c^*0}}^2)\tau/ 2}}]$ and $\tau\approx (0.1-0.2)$ GeV$^{-2}$. We have evaluated the running mass at $\mu=7.5$ GeV.  The result is
comparable to the more involved estimate $f_{B_c}=371(17)$~{\rm MeV} in\,\cite{SNbc20}, 
%%%%%%%%%%%%%%%%%%%%%%%%%%%%%

%%%%%%%%%%%%%%%%%%%%%%%%%%%%%%%%%%%%%%%
\begin{figure}[hbt]
\vspace*{-0.25cm}
\begin{center}
%\centerline {\hspace*{-7.5cm} \bf a) }
\includegraphics[width=11cm]{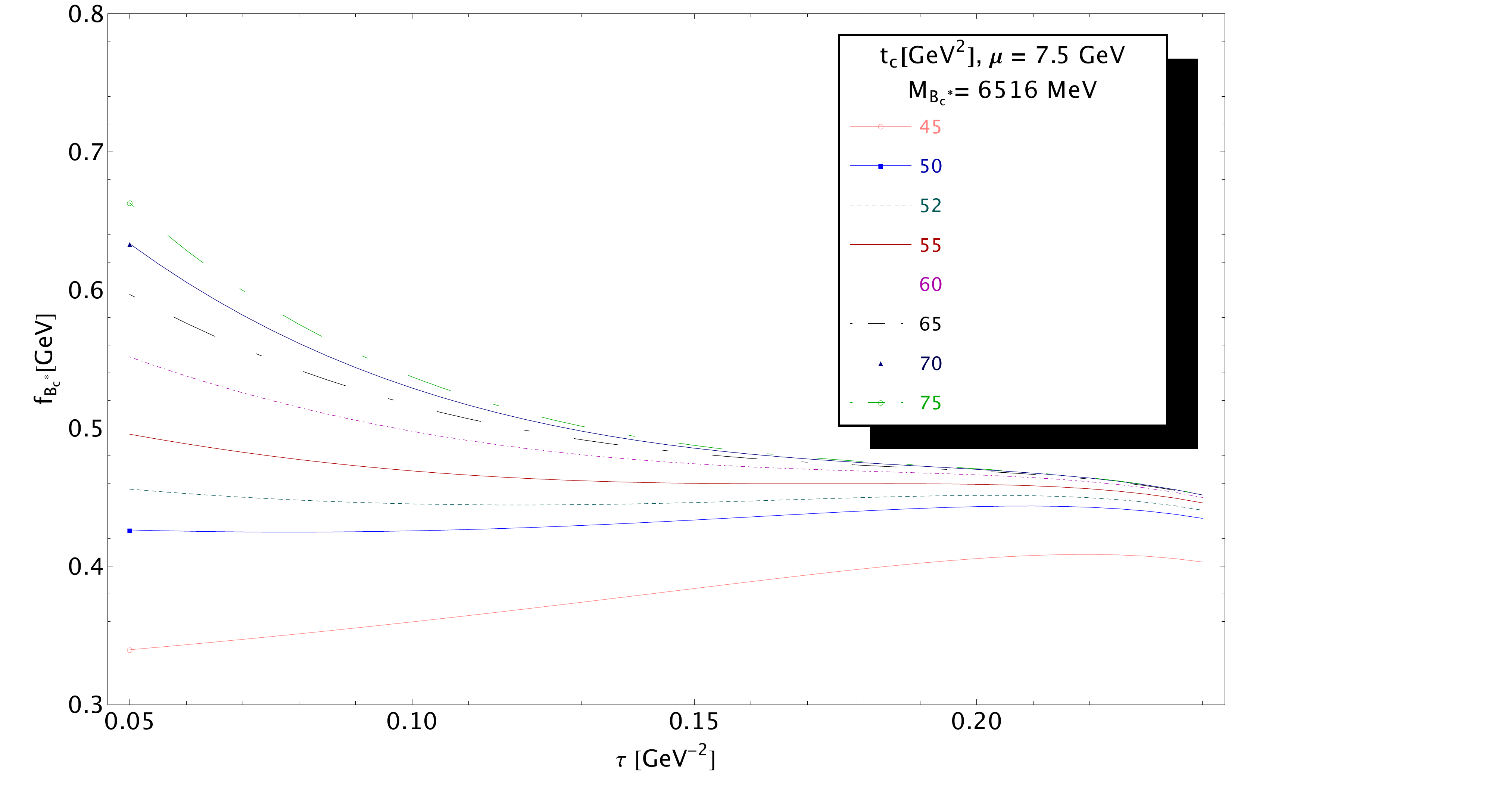}
%\vspace{0.25cm}
%\centerline {\hspace*{-7.5cm} \bf b) }
\includegraphics[width=11.2cm]{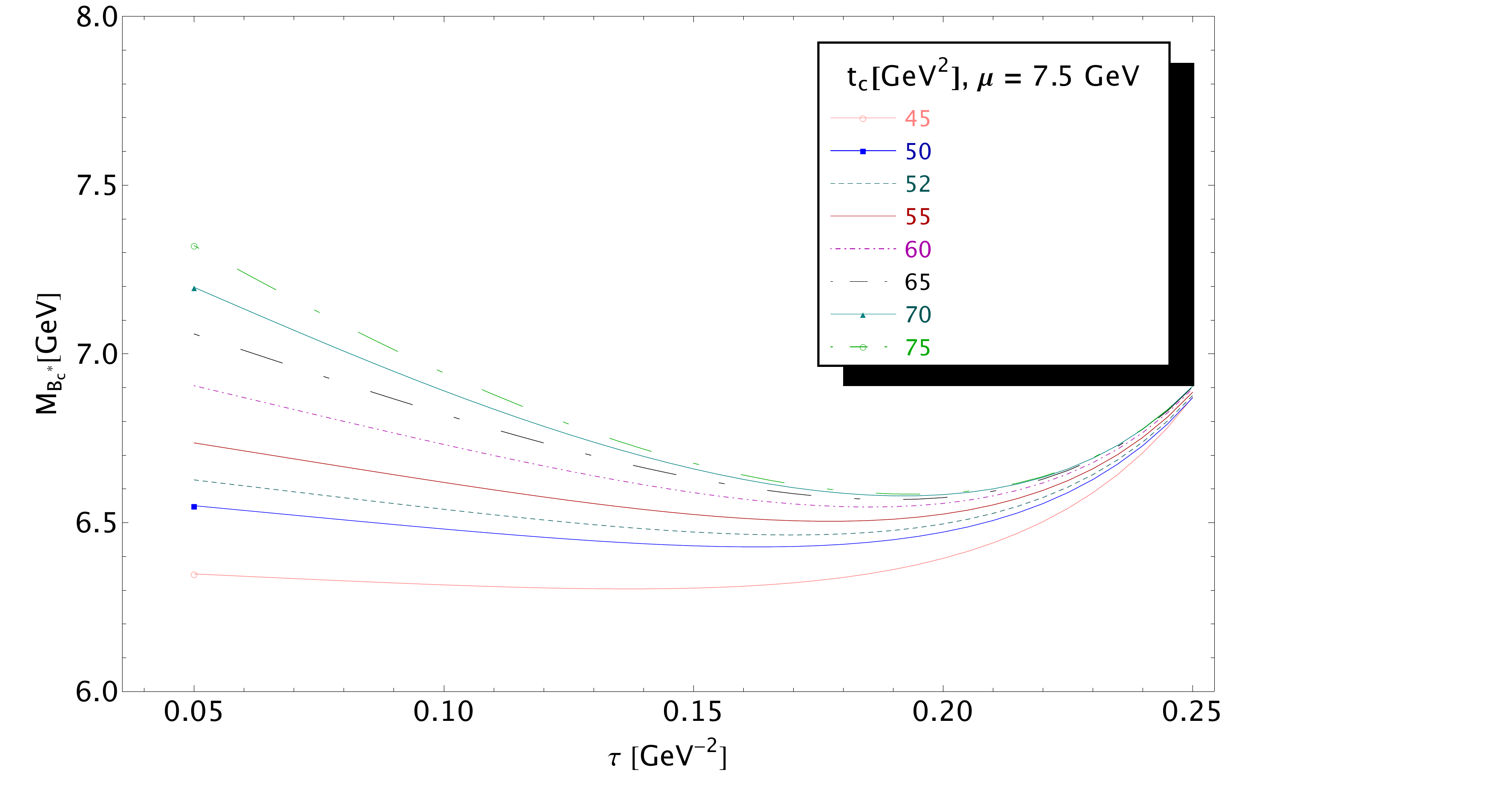}
\vspace*{-0.5cm}
\caption{\footnotesize  $f_{B^*_c}$ and $M_{B^*_c}$ as function of $\tau$ for different values of $t_c$, for $\mu$=7.5 GeV.} 
\label{fig:bcstartau}
\end{center}
\vspace*{-1cm}
\end{figure} 

%%%%%%%%%%%%%%%%%%%%%%%%%%%%%%%%%%%
 %%%%%%%%%%%%%%%%%%%%%%%%%%%%%%%%%%%%%%%
\begin{figure}[hbt]
\vspace*{-0.25cm}
\begin{center}
%\centerline {\hspace*{-7.5cm} \bf a) }
\includegraphics[width=9.5cm]{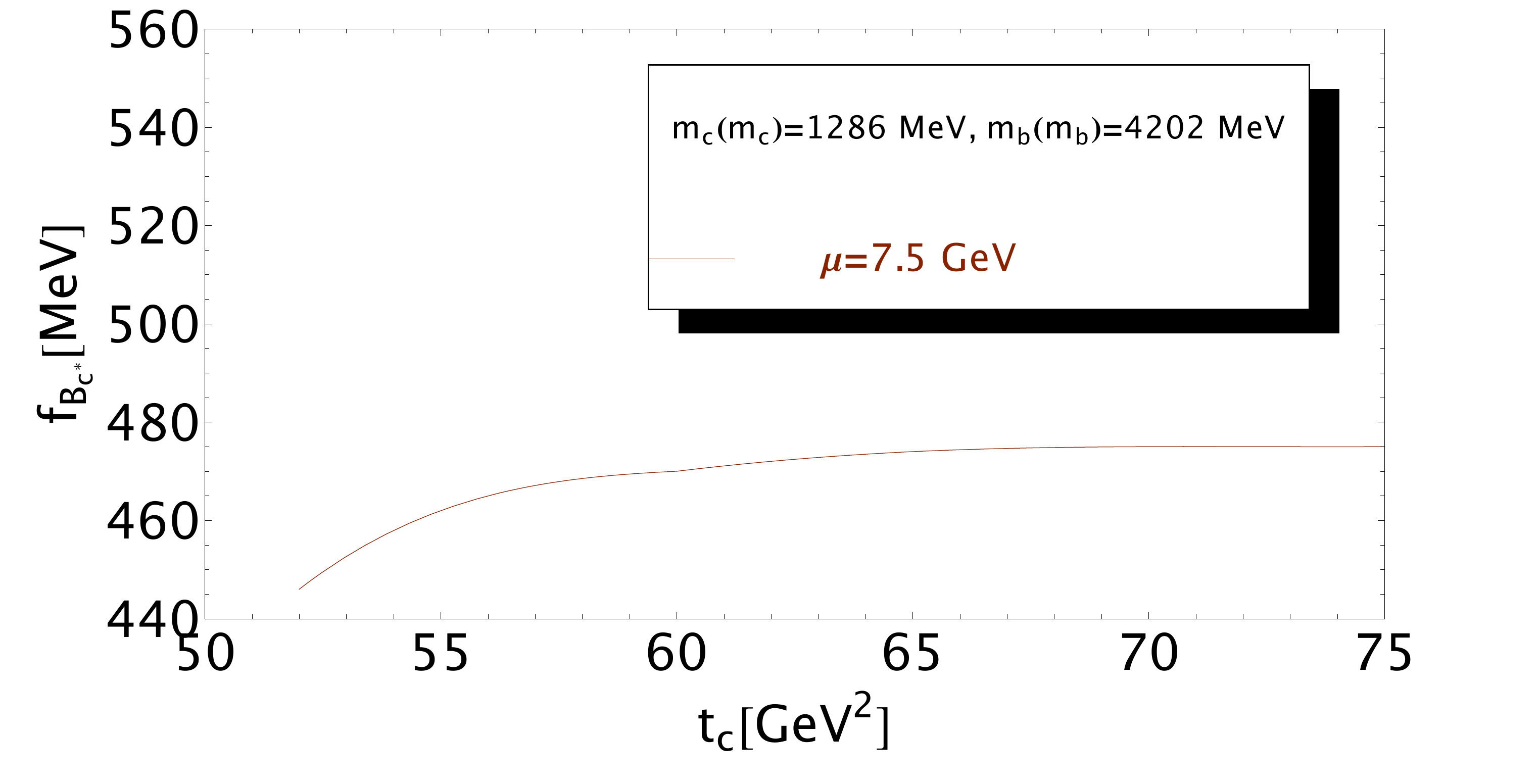}
%\centerline {\hspace*{-7.5cm}\bf b) }\\
%\vspace{0.25cm}
\includegraphics[width=9.5cm]{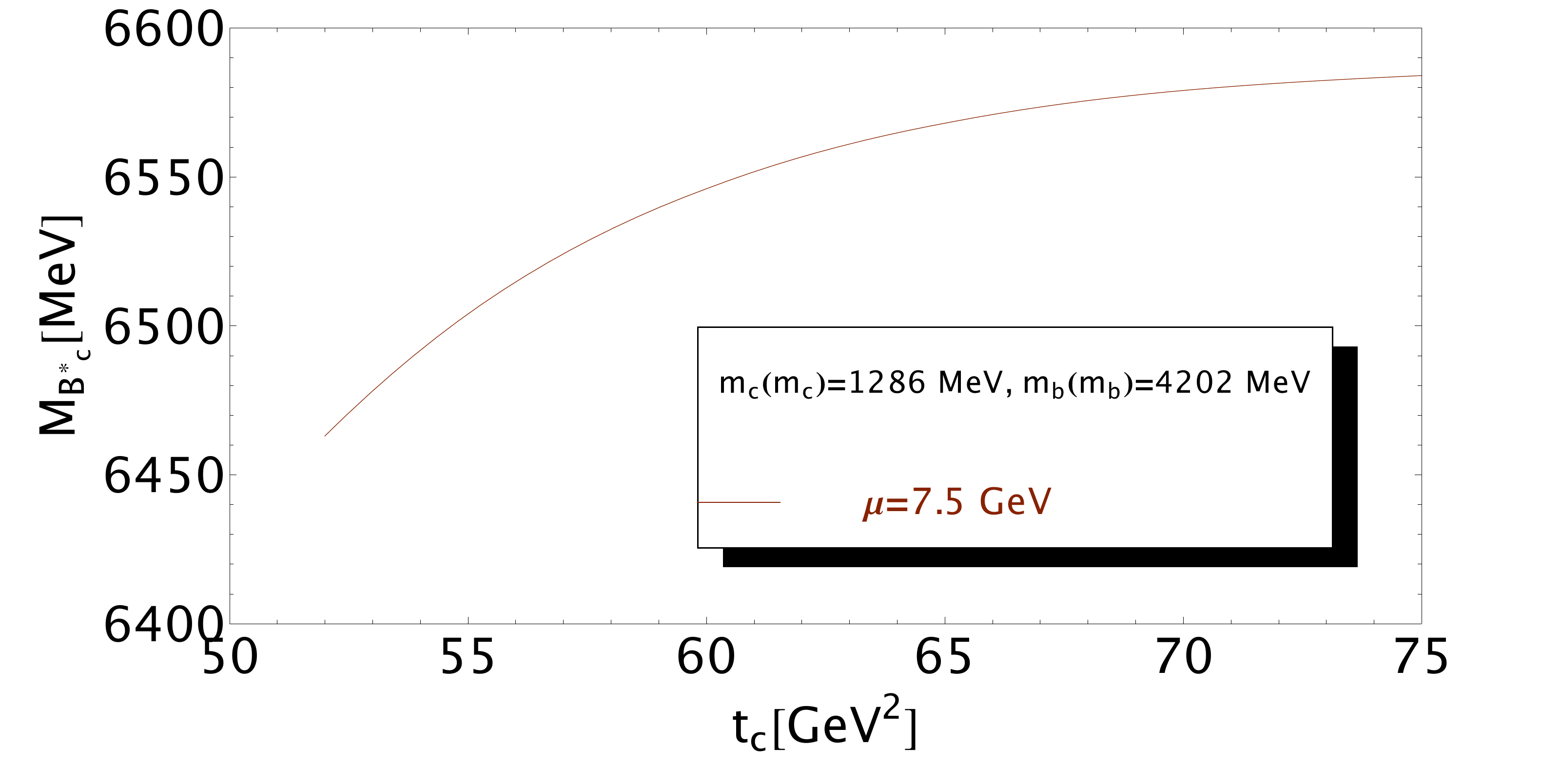}
\vspace*{-0.5cm}
\caption{\footnotesize  $f_{B^*_c}$ and $M_{B^*_c}$ at the inflexion point / minimas of $\tau$ as function of $t_c$ for $\mu$=7.5 GeV.} 
\label{fig:mbcstar-tc}
\end{center}
%\vspace*{-0.75cm}
\end{figure} 
%%%%%%%%%%%%%%%%%%%%%%%%%%%%%%%%%%%%%%%
\begin{figure}[hbt]
\vspace*{-0.25cm}
\begin{center}
%\centerline {\hspace*{-7.5cm} \bf a) }
\includegraphics[width=9.5cm]{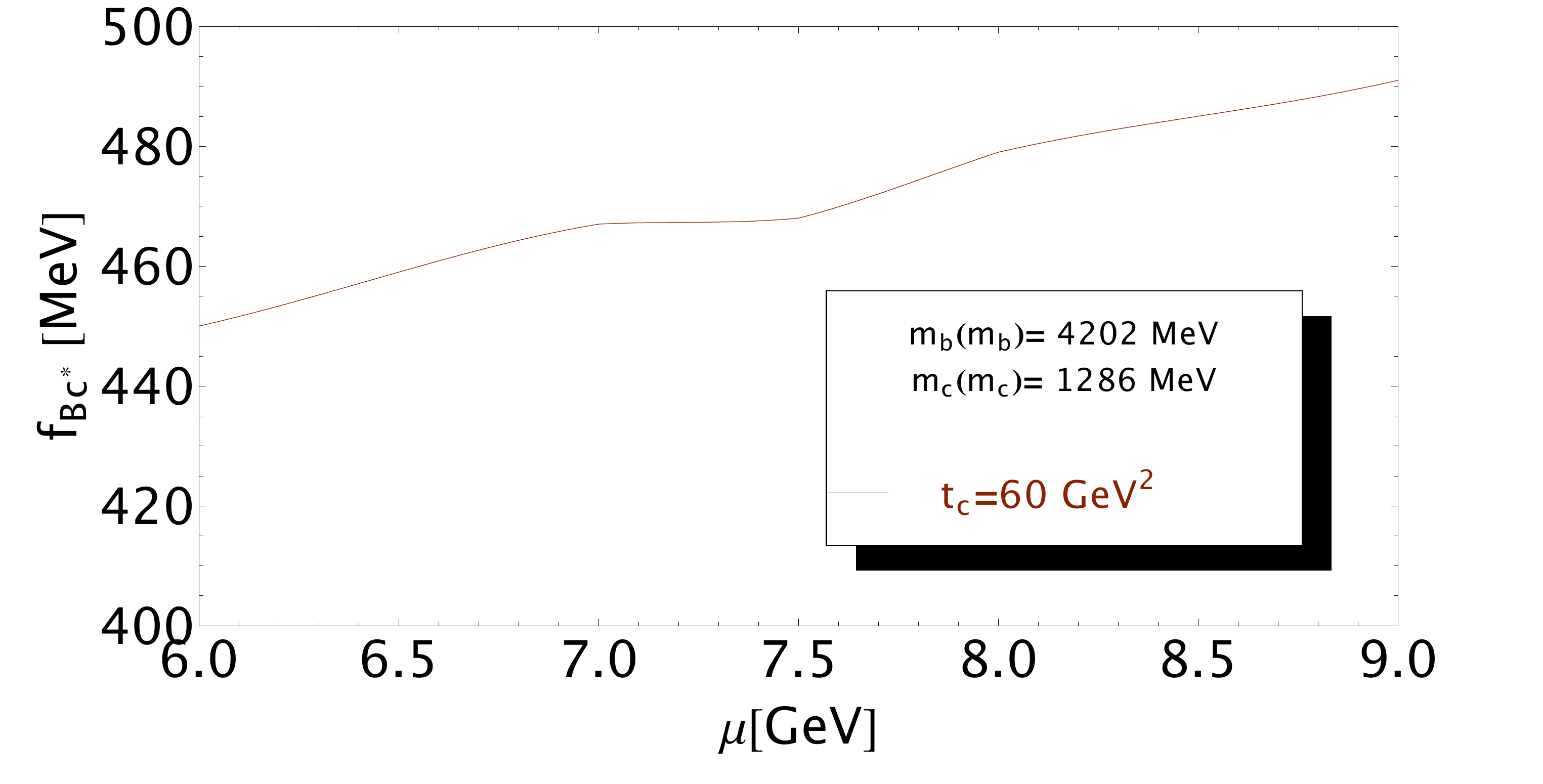}
%\vspace{0.25cm}
%\centerline {\hspace*{-7.5cm} \bf b) }
\includegraphics[width=9.5cm]{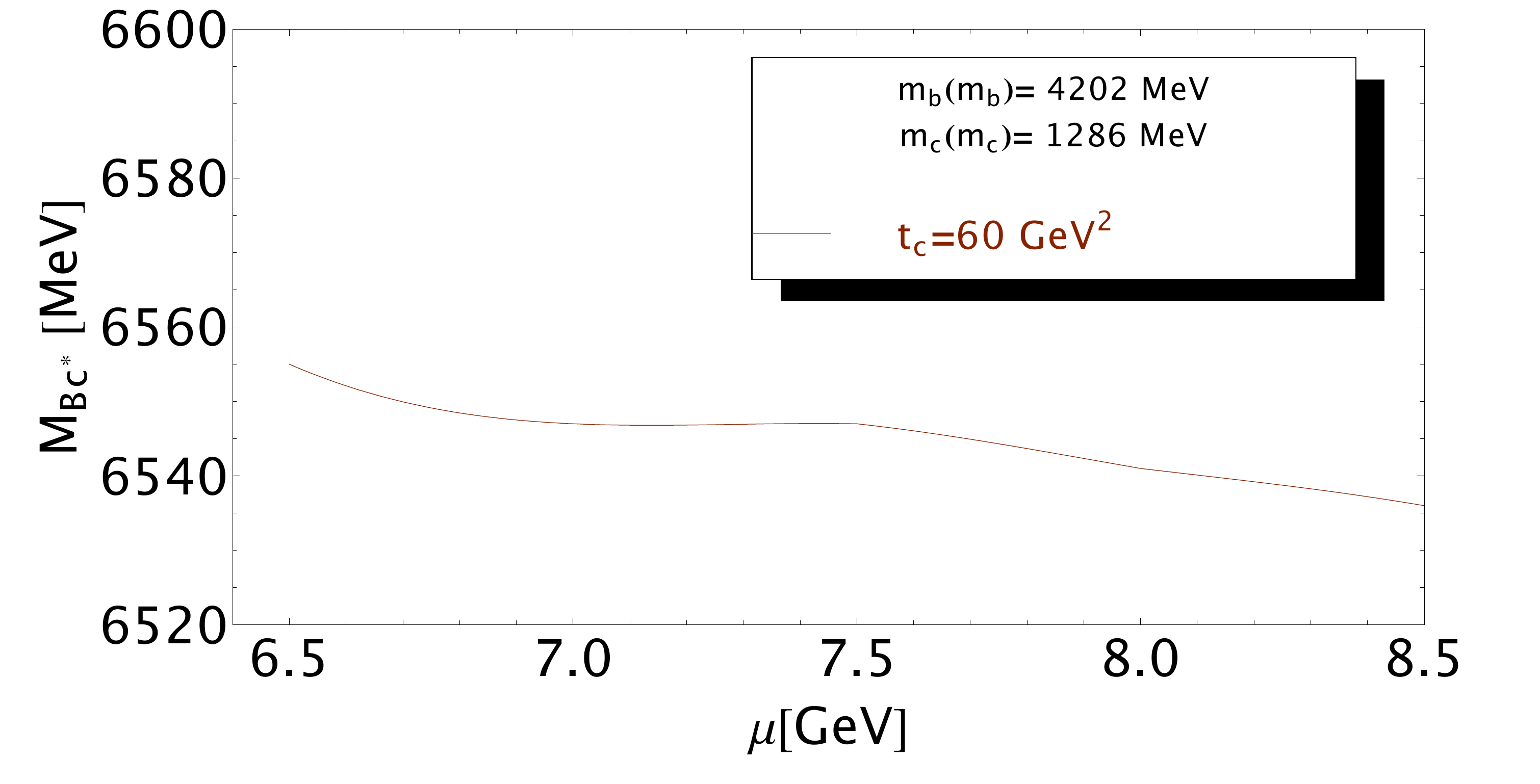}
\vspace*{-0.5cm}
\caption{\footnotesize  $f_{B^*_c}$ and $M_{B^*_c}$ as function of $\mu$ at the $\tau$ inflexion points / minimas.} 
\label{fig:fbcstar-mu}
\end{center}
\vspace*{-0.75cm}
\end{figure} 
%%%%%%%%%%%%%%%%%%%%%%%%%%%%%%%%%%%
 %%%%%%%%%%%%%%%%%%%%%%%%%%%%%
\section{$1^{--}$ Vector channel}
%%%%%%%%%%%%%%%%%%%%%%%%%%%%%
%%%%%%%%%%%%%%%%%%%%%%%%%%%%% 
We do a similar analysis for the vector channel $B^*_c$ which is summarized in the different figures shown below. We think that it is important to present the figures in each channel in order to give a better understanding of the results as the curves have not the same behaviours in $(\tau,t_c,\mu)$.  
%%%%%%%%%%%%%%%%%%%%%%%%%%%%%%%%%%%
\vspace*{-0.25cm}\subsection*{\b  $\tau$-stability}
%%%%%%%%%%%%%%%%%%%%%%%%%%%%%%%%%%%
%%%%%%%%%%%%%%%%%%%%%%%%%%%%%%%%%%%%%%%
%Using the QCD input parameters in Table\,\ref{tab:param}, 
%In a first step, fixing  the value of $\mu=7.5$ GeV which we shall justify later and which is the central value obtained in\,\cite{SNbc20,SNp15}, 
We show in Fig.\,\ref{fig:bcstartau} the $\tau$-behaviour of $f_{B^*_c}$ and $M_{B_c^*}$ for different values of $t_c$.
We see that $f_{B^*_c}$  presents inflexion points and $M_{B^*_c}$  $\tau$-minimas for $t_c\geq$ 52 GeV$^2$. 

%%%%%%%%%%%%%%%%%%%%%%%%%%%%%%%%%%%
\vspace*{-0.25cm}
\vspace*{-0.25cm}\subsection*{\b  $t_c$-stability}
%%%%%%%%%%%%%%%%%%%%%%%%%%%%%%%%%%%
We study the $t_c$-behaviour of $f_{B^*_c}$ and $M_{B^*_c}$ in Fig.\,\ref{fig:mbcstar-tc} where we see that both quantities start to stabilize in $t_c$ for $t_c\simeq 65$ GeV$^2$. 
 %%%%%%%%%%%%%%%%%%%%%%%%%%%%%%%%%%%%%%%

%\vspace*{-0.25cm}
\vspace*{-0.25cm}\subsection*{\b  $\mu$-stability}
%\vspace*{-3cm}
%%%%%%%%%%%%%%%%%%%%%%%%%%%%%%%%%%%
Fixing $t_c=60$ GeV$^2$ , we show in Fig.\,\ref{fig:fbcstar-mu} the $\mu$-behaviour of $M_{B^*_c}$ at the $\tau$-minimas where we note a net inflexion point at :
\beq
\mu \simeq (7\sim 7.5) ~{\rm GeV}~,
%\label{eq:mu}
\eeq
in agreement with the ones obtained in the previous section and quoted in\,\cite{SNp15,SNbc15,SNbc20} indicating the self-consistency of the whole approach. 

%%%%%%%%%%%%%%%%%%%%%%%%%%%%%%%%%%%
\vspace*{-0.25cm}\subsection*{\b  QCD continuum versus lowest resonance contribution}
\begin{figure}[hbt]
%\vspace*{-0.25cm}
\begin{center}
\includegraphics[width=10cm]{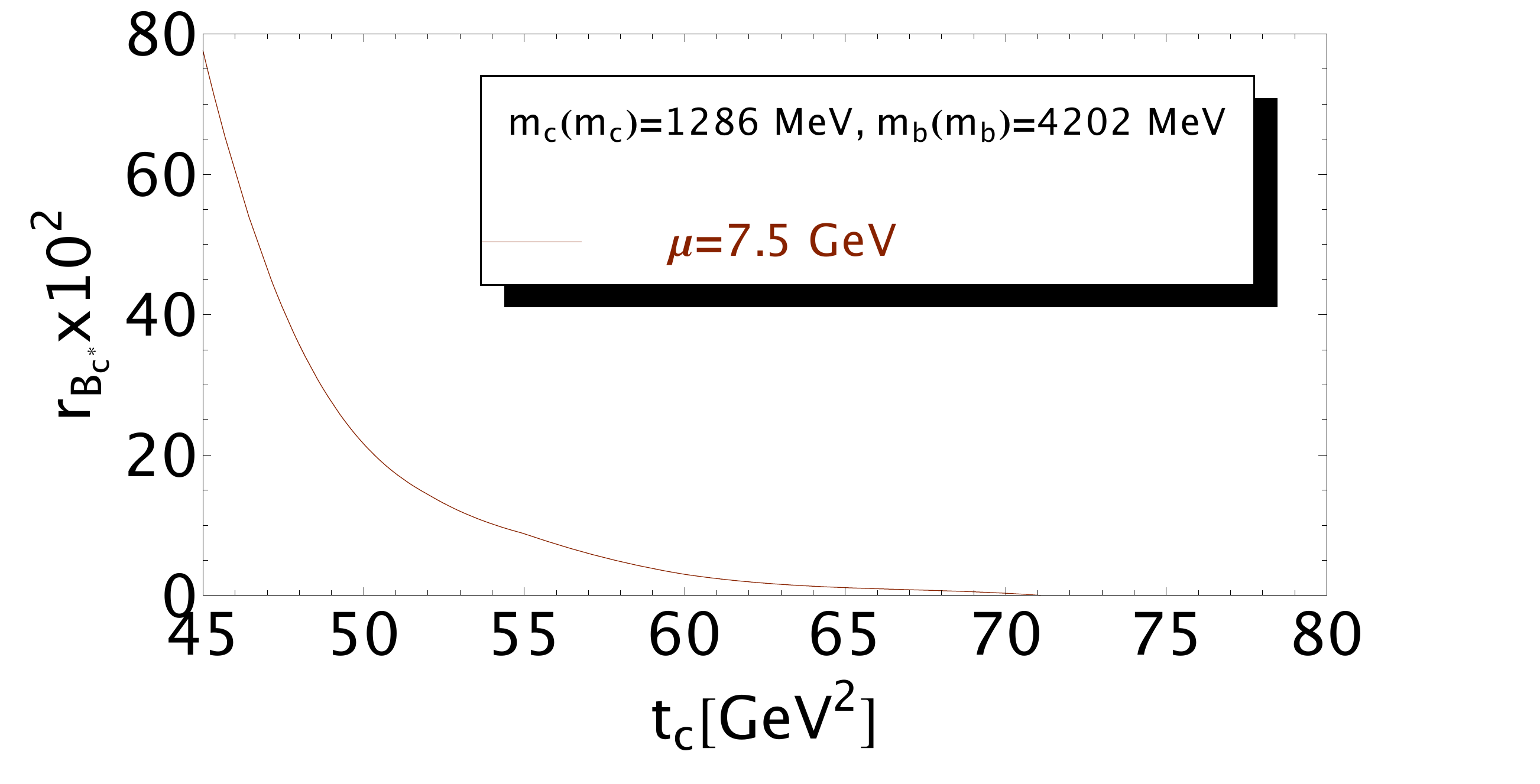}
\vspace*{-0.25cm}
\caption{\footnotesize  Ratio $r_{B^*_c}$ of the continuum over the lowest ground state contribution as function of $t_c$ at the corresponding $\tau$-minimas for $ \mu$=7.5 GeV.} 
\label{fig:rbcstar-cont}
\end{center}
\end{figure}

%%%%%%%%%%%%%%%%%%%%%%%%%%%%%%%%%%%
We show in Fig.\,\ref{fig:rbcstar-cont} the ratio of the continuum over the lowest ground state contribution as predicted by QCD :
\beq
r_{B^*c}\equiv{\int_{t_c}^\infty dt{\rm e}^{-t\tau}{\rm Im} \tilde\Pi^{(1)}_{cont}\over\int_{(m_c+m_b)^2}^{t_c}dt{\rm e}^{-t\tau}{\rm Im}  \tilde\Pi^{(1)}_{B^*_c}}
\eeq
%%%%%%%%%%%%%%%%%%%%%%%%%%%%%%%%%%%%%%%
%%%%%%%%%%%%%%%%%%%%%%%%%%%%%%%%%%%%%%%
%The curve started from $t_c=56.5$ GeV$^2$ where the QCD continuum contribution to the spectral function is half of the resonance contribution
%where one can also note from Fig.\,\ref{fig:bctau} that the $\tau$-stability is reached from this value.
%%%%%%%%%%%%%%%%%%%%%%%%%%%%%%%%%%%
\vspace*{-0.25cm}\subsection*{\b  Predictions for $M_{B^*_c}$ and $f_{B^*_c}$}
%%%%%%%%%%%%%%%%%%%%%%%%%%%%%%%%%%%
From the previous analysis, we take $t_c\simeq (52-65)$ GeV$^2$ for extracting our optimal results. The lowest value of $t_c$ corresponds to the  beginning of $\tau$-stability and also here to the ``QCD continuum" contribution which is less than 20\% of the resonance one. The highest value corresponds to the beginning of $t_c$-stability. 
We obtain in units of MeV:
\bea
 \hspace*{-0.25cm}M_{B^*_c}  \hspace*{-0.cm}&\simeq&\hspace*{-0.cm} 6451(52)_{t_c}(1)_\tau (1)_\mu (11)_{m_{b,c}}(7)_{\alpha_s}(17)_{G^2}(65)_{syst},\nnb\\
  \hspace*{-0.25cm}f_{B^*_c}\hspace*{-0.cm}&\simeq&\hspace*{-0.cm} 442(41)_{M_{B_c^*}}(11)_{t_c}(1)_\mu(1)_\tau(6)_{m_{b,c}}(7)_{\alpha_s}(4)_{G^2}%(31)_{syst}
  \,.
  \eea
 
%  where one can notice that the uncertainites due to the QCD parameters are realtively small. These results are collected in Table\,\ref{tab:result} and compared to the ones from some other approaches.

 %%%%%%%%%%%%%%%%%%%%%%%%%%%%%
\section{$1^{++}$ Axial-Vector channel}
%%%%%%%%%%%%%%%%%%%%%%%%%%%%% 
We do a similar analysis for the $B_{c1}$ axial-vector meson. The expression of the two-point function can be deduced from the vector one by changing $m_c$ to $-m_c$. The anaysis is summarized through the  different figures shown below. 
%%%%%%%%%%%%%%%%%%%%%%%%%%%%%%%%%%%
\vspace*{-0.25cm}\subsection*{\b  $\tau$-stability}
%%%%%%%%%%%%%%%%%%%%%%%%%%%%%%%%%%%
 %%%%%%%%%%%%%%%%%%%%%%%%%%%%%%%%%%%%%%%
\begin{figure}[hbt]
\vspace*{-0.25cm}
\begin{center}
%\centerline {\hspace*{-7.5cm} \bf a) }
\includegraphics[width=11cm]{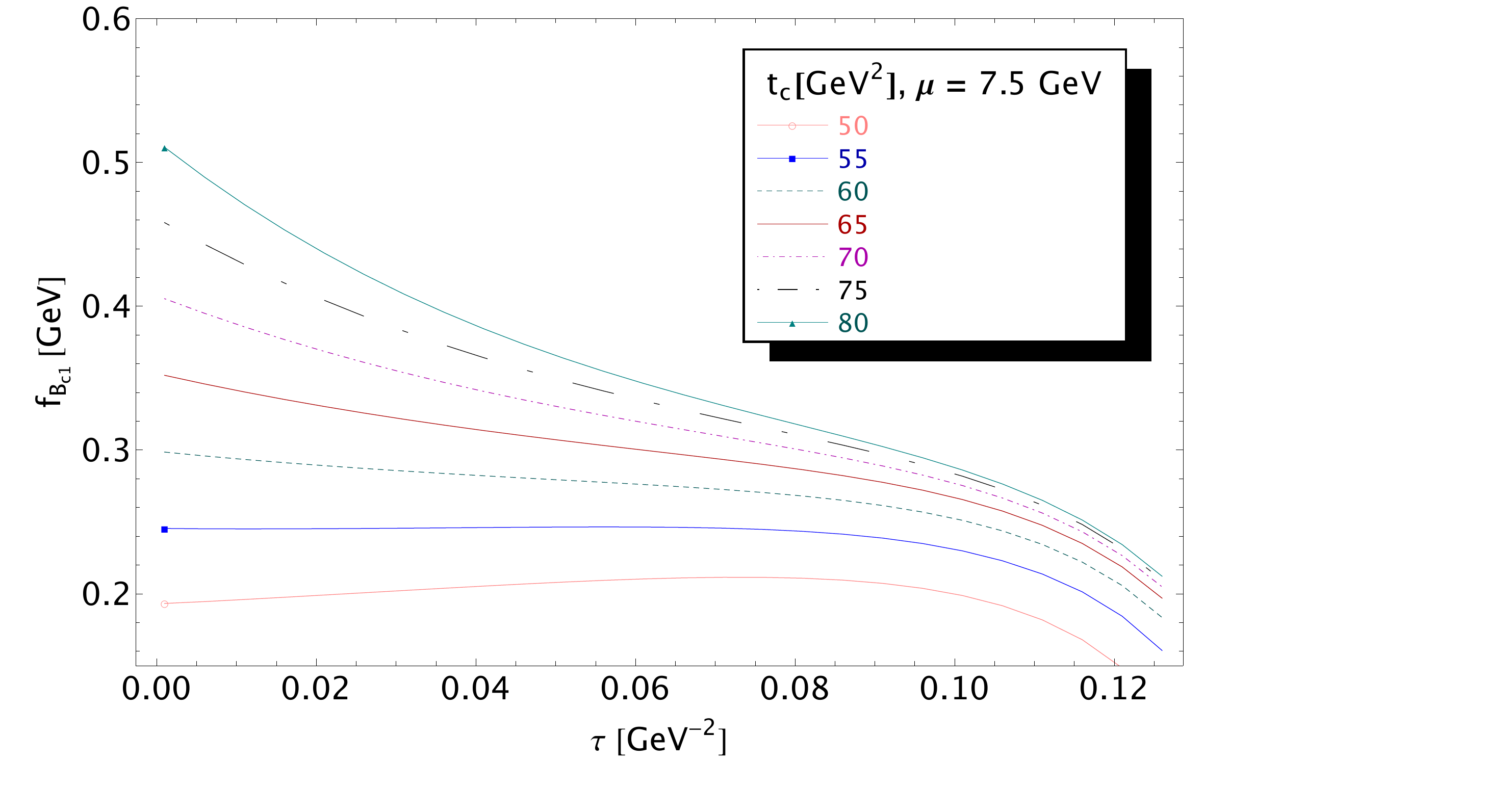}
%\vspace{0.25cm}
%\centerline {\hspace*{-7.5cm} \bf b) }
\includegraphics[width=11.2cm]{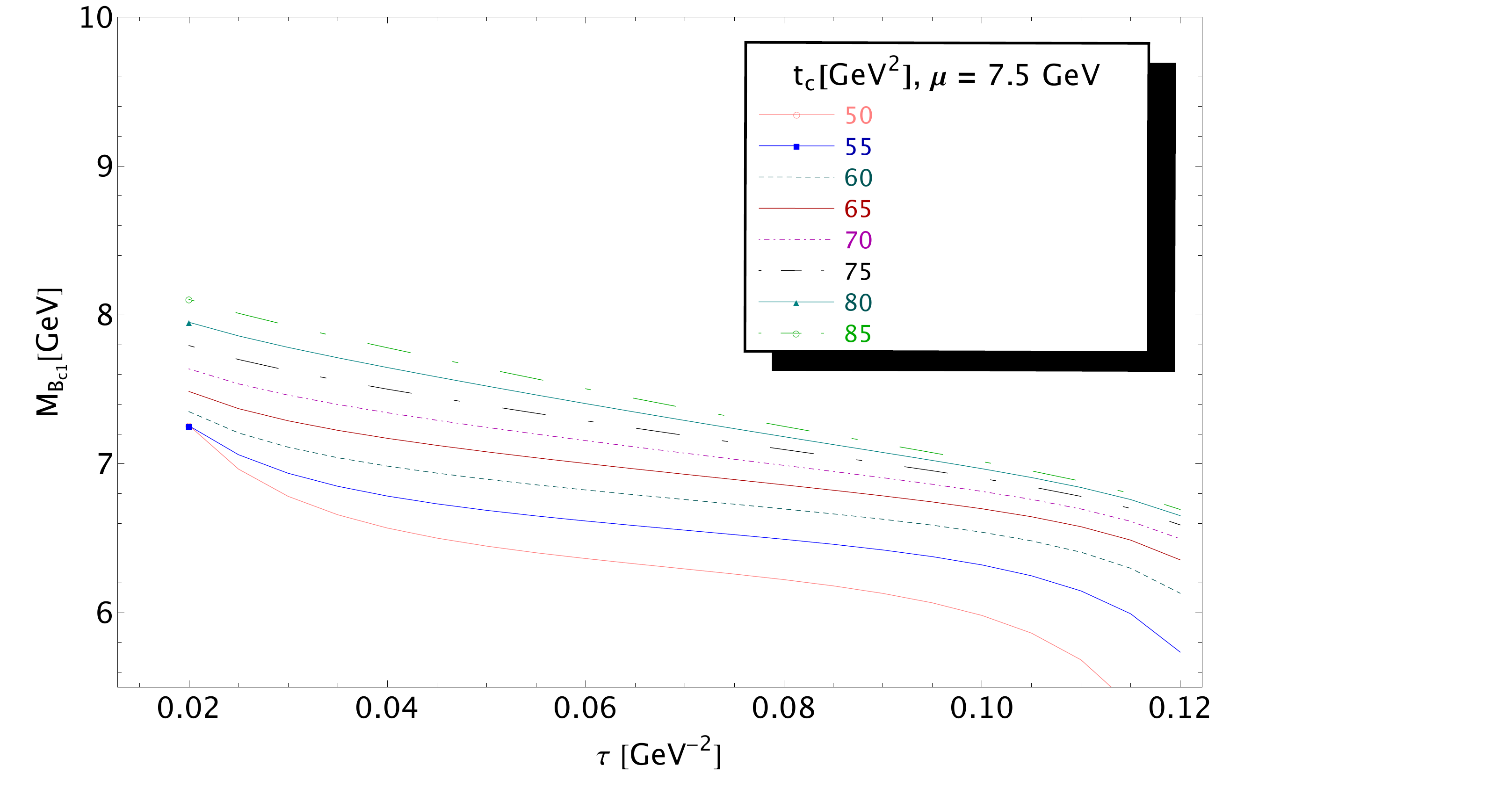}
\vspace*{-0.5cm}
\caption{\footnotesize  $f_{B_{c1}}$ and $M_{B_{c1}}$ as function of $\tau$ for different values of $t_c$ and for $\mu$=7.5 GeV.} 
\label{fig:bca-tau}
\end{center}
%\vspace*{-0.75cm}
\end{figure} 
%%%%%%%%%%%%%%%%%%%%%%%%%%%%%%%%%%%%%%%
%Using the QCD input parameters in Table\,\ref{tab:param}, 
%In a first step, fixing  the value of $\mu=7.5$ GeV which we shall justify later and which is the central value obtained in\,\cite{SNbc20,SNp15}, 
We show in Fig.\,\ref{fig:bca-tau} the $\tau$-behaviour of $f_{B_{c1}}$ and $M_{B_{c1}}$ for different values of $t_c$.
We see that both quantities present inflexion points for $\tau\simeq (0.09-0.10)$ GeV$^{-2}$ which appear for $t_c\geq$ 50 GeV$^2$. Imposing that the ``QCD continuum" contribution is less than 20-25\% of the resonance one leads to  $t_c\geq$ 65 GeV$^2$. 
%%%%%%%%%%%%%%%%%%%%%%%%%%%%%%%%%%%
\vspace*{-0.25cm}\subsection*{\b  $t_c$-stability}
\begin{figure}[hbt]
\vspace*{-0.25cm}
\begin{center}
%\centerline {\hspace*{-7.5cm} \bf a) }
\includegraphics[width=9.5cm]{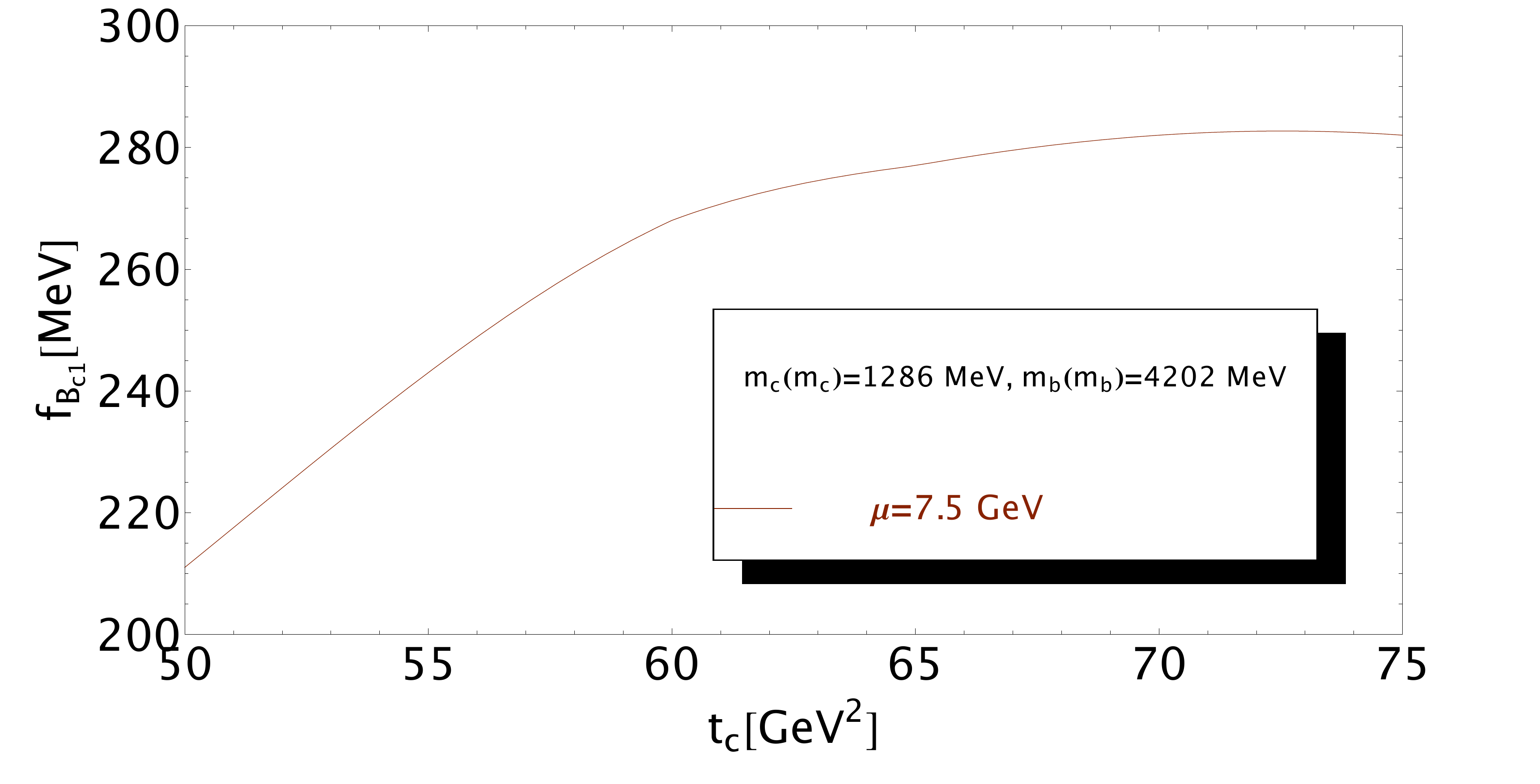}
%\centerline {\hspace*{-7.5cm}\bf b) }\\
%\vspace{0.25cm}
%\includegraphics[width=9.5cm]{mbca-tc.pdf}
\vspace*{-0.5cm}
\caption{\footnotesize  $f_{B_{c1}}$ at the inflexion point of $\tau$ as function of $t_c$ for $\mu$=7.5 GeV.} 
\label{fig:fbca-tc}
\end{center}
%\vspace*{-0.75cm}
\end{figure}

\begin{figure}[hbt]
\vspace*{-0.25cm}
\begin{center}
\includegraphics[width=9.5cm]{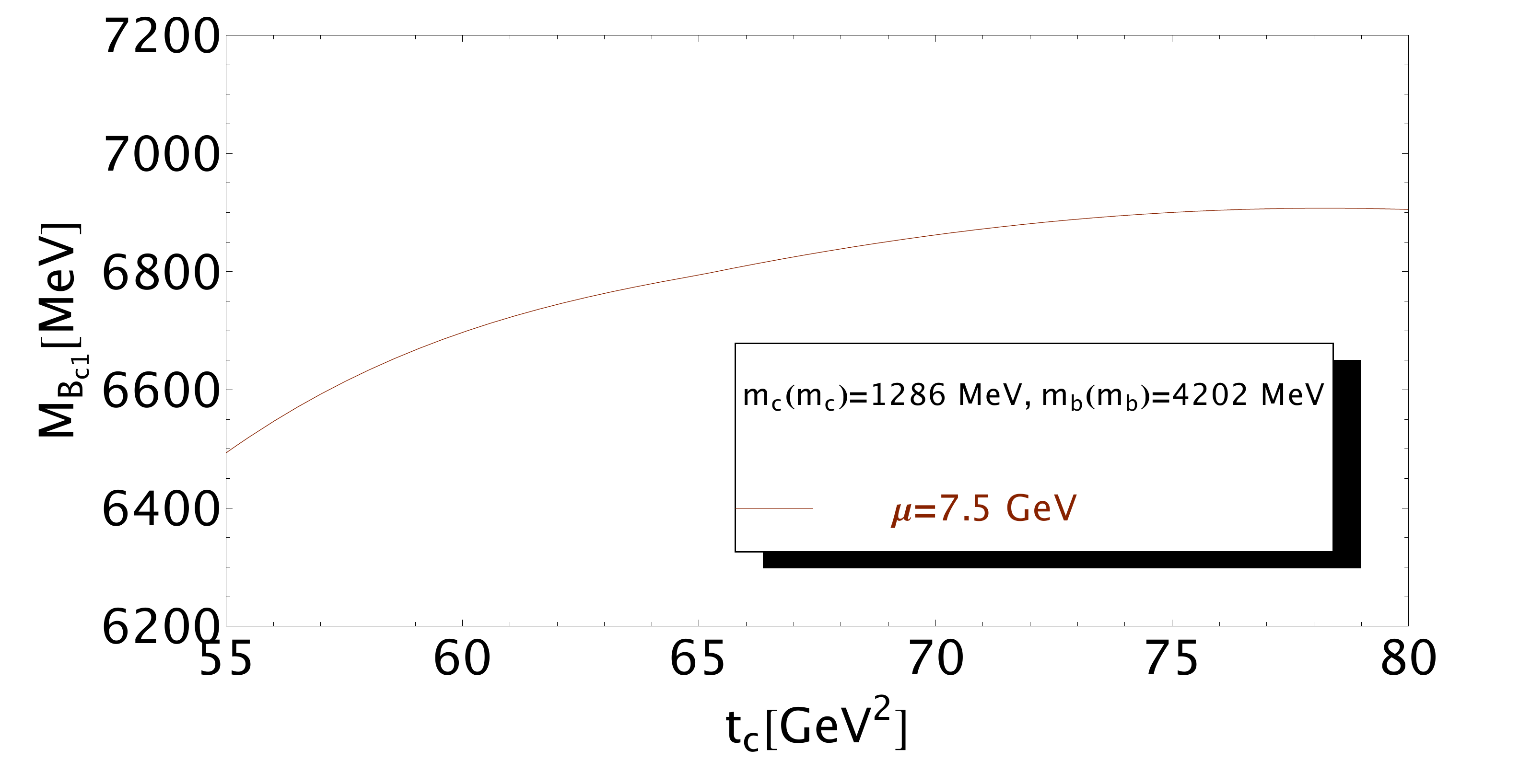}
\vspace*{-0.5cm}
\caption{\footnotesize  $M_{B_{c1}}$ at the inflexion point of $\tau$ as function of $t_c$ for $\mu$=7.5 GeV.} 
\label{fig:mbca-tc}
\end{center}
%\vspace*{-0.75cm}
\end{figure}

%%%%%%%%%%%%%%%%%%%%%%%%%%%%%%%%%%%
We study the $t_c$-behaviour of $f_{B_{c1}}$ and $M_{B_{c1}}$ in Figs. \,\ref{fig:fbca-tc} and \,\ref{fig:mbca-tc} where both quantities start to stabilize for $t_c\simeq 65$ GeV$^2$.  The beginning of $t_c$-stability is reached for $t_c\approx 75$ GeV$^2$. For definiteness, we shall work in the range of $t_c\simeq (65-75)$   GeV$^2$.
%%%%%%%%%%%%%%%%%%%%%%%%%%%%%%%%%%%%%%%
%%%%%%%%%%%%%%%%%%%%%%%%%%%%%%%%%%%%%%%
%%%%%%%%%%%%%%%%%%%%%%%%%%%%%%%%%%%%%%%
%%%%%%%%%%%%%%%%%%%%%%%%%%%%%%%%%%%%%%%
 %%%%%%%%%%%%%%%%%%%%%%%%%%%%%%%%%%%%%%%
\begin{figure}[hbt]
\vspace*{-0.25cm}
\begin{center}
%\centerline {\hspace*{-7.5cm} \bf a) }
\includegraphics[width=9.5cm]{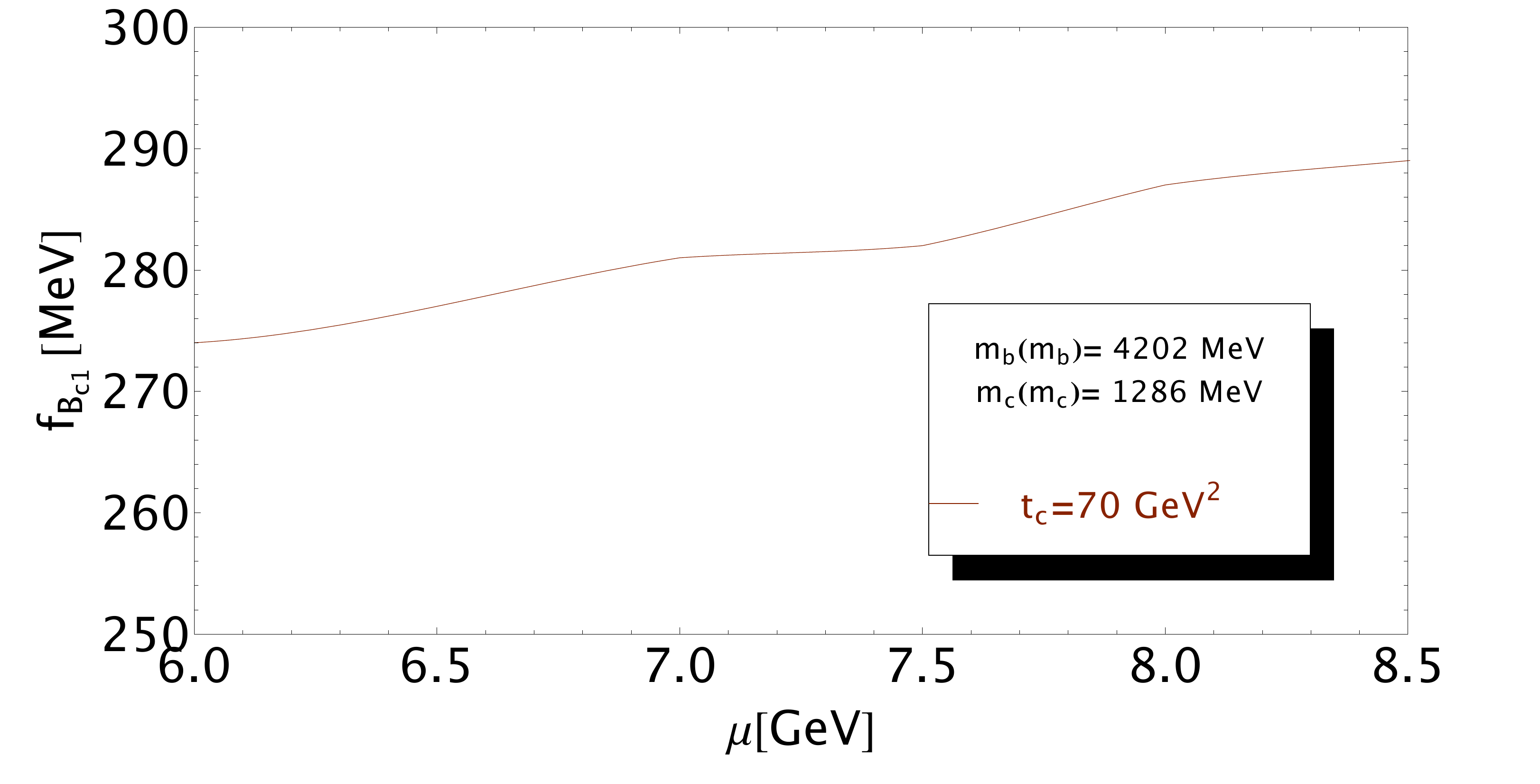}
%\vspace{0.25cm}
%\centerline {\hspace*{-7.5cm} \bf b) }
%\includegraphics[width=9.5cm]{mbca-mu.pdf}
\vspace*{-0.5cm}
\caption{\footnotesize  $f_{B_{c1}}$ as function of $\mu$ for $\tau\simeq 0.095$ GeV$^{-2}$.} 
\label{fig:fbca-mu}
\end{center}
%\vspace*{-0.75cm}
\end{figure} 

%%%%%%%%%%%%%%%%%%%%%%%%%%%%%%%%%%%%%%%
\begin{figure}[hbt]
\vspace*{-0.25cm}
\begin{center}
%\centerline {\hspace*{-7.5cm} \bf a) }
%\includegraphics[width=9.5cm]{fbca-mu.pdf}
%\vspace{0.25cm}
%\centerline {\hspace*{-7.5cm} \bf b) }
\includegraphics[width=9.5cm]{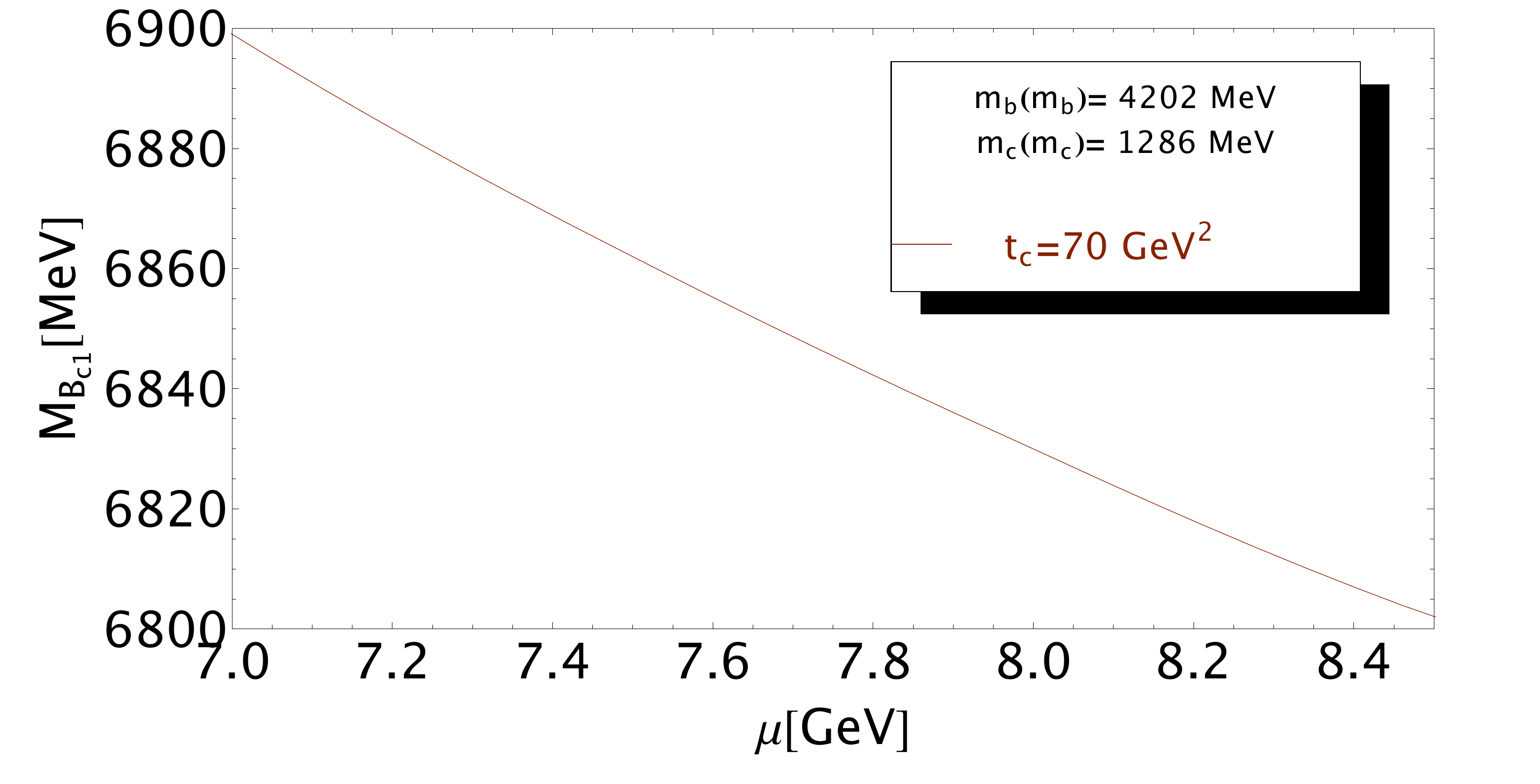}
\vspace*{-0.5cm}
\caption{\footnotesize  $M_{B_{c1}}$ as function of $\mu$ for $\tau\simeq 0.095$ GeV$^{-2}$.} 
\label{fig:mbca-mu}
\end{center}
\vspace*{-0.5cm}
\end{figure} 

 %%%%%%%%%%%%%%%%%%%%%%%%%%%%%%%%%%%
\vspace*{-0.25cm}\subsection*{\b  $\mu$-stability}
%%%%%%%%%%%%%%%%%%%%%%%%%%%%%%%%%%%
Fixing $t_c=70$ GeV$^2$ , we show in Figs.\,\ref{fig:fbca-mu} and \,\ref{fig:mbca-mu} the $\mu$-behaviour of  $f_{B_{c1}}$ and $M_{B_{c1}}$ for $\tau\simeq 0.095$ GeV$^{-2}$.  We note that $M_{B_{c1}}$ is a decreasing function of $\mu$ while $f_{B_{c1}}$ presents a net inflexion point at :
\beq
\mu \simeq (7\sim 7.5) ~{\rm GeV}~.
%\label{eq:mu}
\eeq
This value agrees with the ones obtained in the previous sections and in\,\cite{SNp15,SNbc20} showing again the self-consistency of the whole approach for the $B_c$-like mesons. 
%%%%%%%%%%%%%%%%%%%%%%%%%%%%%%%%%%%
\vspace*{-0.25cm}\subsection*{\b  QCD continuum versus lowest resonance contribution}
\begin{figure}[hbt]
\vspace*{-0.25cm}
\begin{center}
\includegraphics[width=10cm]{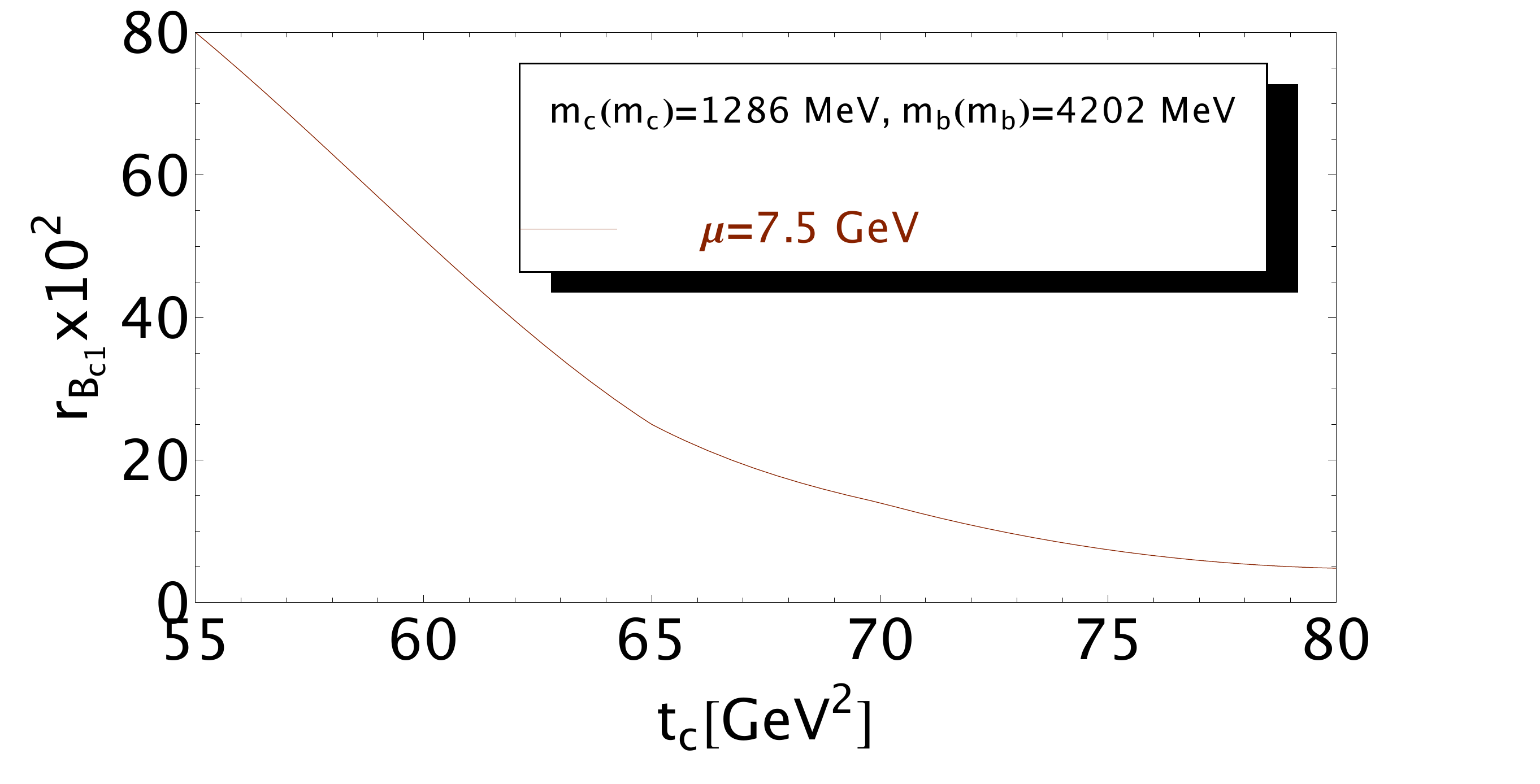}
\vspace*{-0.5cm}
\caption{\footnotesize  Ratio $r_{B_{c1}}$ of the continuum over the lowest ground state contribution as function of $t_c$ at the corresponding $\tau$-inflexion point for $ \mu$=7.5 GeV.} 
\label{fig:rbca-cont}
\end{center}
\vspace*{-0.5cm}
\end{figure}

%%%%%%%%%%%%%%%%%%%%%%%%%%%%%%%%%%%
We show in Fig.\,\ref{fig:rbca-cont} the ratio of the continuum over the lowest ground state contribution as predicted by QCD:
%\beq
%r_{B^*c}\equiv{\int_{t_c}^\infty dt{\rm e}^{-t\tau}{\rm Im} \tilde\Pi^{(1)}_{cont}\over\int_{(m_c+m_b)^2}^{t_c}dt{\rm e}^{-t\tau}{\rm Im}  \tilde\Pi^{(1)}_{B_{c1}}}
%\eeq
%%%%%%%%%%%%%%%%%%%%%%%%%%%%%%%%%%%%%%%
%%%%%%%%%%%%%%%%%%%%%%%%%%%%%%%%%%%%%%%
%The curve started from $t_c=56.5$ GeV$^2$ where the QCD continuum contribution to the spectral function is half of the resonance contribution
%where one can also note from Fig.\,\ref{fig:bctau} that the $\tau$-stability is reached from this value.
%%%%%%%%%%%%%%%%%%%%%%%%%%%%%%%%%%%
\vspace*{-0.25cm}\subsection*{\b  Predictions for $M_{B_{c1}}$ and $f_{B_{c1}}$}
%%%%%%%%%%%%%%%%%%%%%%%%%%%%%%%%%%%
From the previous analysis we take $t_c\simeq (65-75)$ GeV$^2$ for extracting our optimal results. The lowest value of $t_c$ corresponds to the  case where the QCD contribution is less than 25\% of the resonance one. The highest value corresponds to the beginning of $t_c$-stability. 
We obtain in units of MeV :
\bea
 \hspace*{-0.1cm}M_{B_{c1}}\hspace*{-0.cm}&\simeq&\hspace*{-0.cm} 6794(68)_{t_c}(44)_\tau (32)_\mu (16)_{m_{b,c}}(60)_{\alpha_s}(11)_{G^2}(68)_{syst}\,,\nnb\\
  \hspace*{-0.25cm}f_{B_{c1}} \hspace*{-0.cm}&\simeq&\hspace*{-0.cm}274(19)_{M_{B_{c1}}}(10)_{t_c}(1)_\mu(7)_\tau(1)_{m_{b,c}}(3)_{\alpha_s}(1)_{G^2}~. %(10)_{syst}~.
  \eea
   %%%%%%%%%%%%%%%%%%%%%%%%%%%%%%%%%%%%%%%
\section{Comments on the results} \label{sec:results}
%%%%%%%%%%%%%%%%%%%%%%%%%%%%%%%%%%%%%%%%
%\vspace*{-0.25cm}\subsection*{\b  Comments on the results}
%%%%%%%%%%%%%%%%%%%%%%%%%%%%%%%%%%%%%%%%%
We notice from previous analysis that :

--  The results from different channels stabilize at a common value of $\mu$ around 7.5 GeV which is consistent with previous analysis in\,\cite{SNp15,SNbc20} indicating the self-consistency of the whole approach.

-- The value of $t_c\simeq (60\pm 5)$ GeV$^2$ where the $B^*_c$ parameters are optimally extracted are about the same as the one of $B_c$ but lower than the ones $t_c\simeq (70\pm 5)$ GeV$^2$ where the $B^*_{0c}$ and $B^*_{c1}$ sum rules are optimized. This feature is dual to the low masses of $(B_c,B^*_c)$ compared to  the ones of $(B^*_{0c},B^*_{c1})$. 

-- The errors due to the QCD parameters are relatively small. The ones from the sum rule external parameters ($t_c,\tau$) are dominant.  In addition to these, the ones on the decay constants are strongly affected by the error on the mass determination.  

-- As mentioned earlier, we have added the systematic error of 1\% on the mass determination and divided the prediction by 1.01 for quantifying the approximate expression expanded in terms of $m_c$ for the non-perturbative contributions. 
%%%%%%%%%%%%%%%%%%%%%%%%%%%%%%%%%%%%%%%%
\section{Mass-splittings from Heavy Quark Symmetry (HQS)}
%%%%%%%%%%%%%%%%%%%%%%%%%%%%%%%%%%%%%%%%%
To improve the predictions on the meson masses, we shall use the properties of Heavy Quark Symmetry (HQS)\,\cite{ISGUR,NEU}. 
In so doing, we confront the observed values of the $D_{(s)},B_{(s)}$-like mass-splittings to the LO expectations of HQS in the heavy quark $(1/M_Q)$ inverse mass expansion where : $Q\equiv c,b$. Then,  we extrapolate this result for predicting the $B_c$-like meson masses. 
%%%%%%%%%%%%%%%%%%%
\vspace*{-0.25cm}\subsection*{\b  Hyperfine splittings}
%%%%%%%%%%%%%%%%%%%
From spin symmetry, one expects that, to LO, the hyperfine splittings are independent on the flavour of the ``brown muck"\,\cite{ISGUR,NEU} which is realized 
%for the observed mass-splittings
experimentally\,\cite{PDG}. In units of GeV$^2$, one has indeed :
\bea
 \hspace*{-0.5cm}M_{B^*}^2-M_{B}^2& \hspace*{-0.cm}=\hspace*{-0.cm}& 0.488\approx M_{D^*}^2-M_{D}^2=0.543~, \nnb\\
 \hspace*{-0.5cm}M_{B^*_s}^2-M_{B_s}^2& \hspace*{-0.cm}=\hspace*{-0.cm}&0.518\approx M_{D^*_s}^2-M_{D_s}^2=0.588~,
\eea
with neglible errors. These results indicate that the $1/M_Q$ corrections to LO are quite small. We shall use the $B$ meson results, where the $1/M_b$ corrections are smaller than the one of the $D$-mesons. Extrapolating to the $B_c$-like mesons and using $M_{B_c}$= 6274.9(0.8) MeV, one can deduce :
\bea
 \hspace*{-0.5cm}M_{B^*_c} &\simeq & (6315\pm 1)~{\rm MeV}~.
 \eea
 %%%%%%%%%%%%%%%%%%%
\vspace*{-0.25cm}\subsection*{\b  Heavy Flavour Independence of Excitation Energies}
%%%%%%%%%%%%%%%%%%%
-- One also expects from HQS that the excitation energies for states with different quantum numbers of the light degrees of freedom are heavy flavour independent\,\cite{NEU,ISGUR},  which is approximately reproduced by the 
data\,\cite{PDG}. In units of MeV, we have :
\bea
 \hspace*{-0.cm}M_{B_1}-M_{B}& \hspace*{-0.cm}=\hspace*{-0.cm}&447(1)\approx M_{D_1}-M_{D}=551(1), \nnb\\
  \hspace*{-0.cm}M_{B_{s1}}-M_{B_s}& \hspace*{-0.cm}=\hspace*{-0.cm}& 462(1)\approx M_{D_{s1}}-M_{D_s}=491(1)
  \eea
  and : 
  \bea
 \hspace*{-0.cm}M_{B_2}-M_{B}& \hspace*{-0.cm}=\hspace*{-0.cm}&458(1)\approx M_{D_2}-M_{D}=596(1),\nnb\\
  \hspace*{-0.cm}M_{B_{s2}}-M_{B_s}& \hspace*{-0.cm}=\hspace*{-0.cm}& 473(0)\approx M_{D_{s2}}-M_{D_s}=601(1). 
 \eea
The approximate equalities between the $B$ and $D$ mass-splittings again indicate that the $1/M_Q$ corrections to the LO relations are negligible. Extrapolating the values for the $B$  to the $B_c$-like mesons, we deduce in units of MeV:
 \bea
 \hspace*{-0.cm} M_{B_{c1}} = (6730\pm 8)~~~{\rm and}~~~ M_{B_{c2}} = (6741\pm 8)~.
\eea
-- For estimating the scalar meson mass $M_{B^*_{0,c}}$, we assume the flavour independence (within the errors)  of the mass-splitting 
of chiral multiplets as given by the sum rule results\,\cite{SNB1,SNBc,SNhl05}:
\beq
  \hspace*{-0.5cm} M_{B^*_{0}}-M_{B}\simeq 422(196)~{\rm MeV}
  \approx M_{D^*_{s0}}-M_{D_s}%\simeq 328(113),
\eeq
and the data\,\cite{PDG}:
\beq
  \hspace*{-0.5cm} M_{D^*_{0}}-M_{D}\simeq 448(29)~{\rm MeV}.
  %\approx M_{D^*_{s0}}-M_{D_s}\simeq 349~{\rm MeV}.
\eeq
Using the previous experimental value,
% to get the $B^*_0$ and $B^*_{0c}$ mesons
we deduce :
\beq
 \hspace*{-0.cm}M_{B^*_0}\simeq 5733~{\rm MeV}~{\rm and}~ M_{B^*_{c0}}\simeq 6723~{\rm MeV}. 
\eeq
The value of $M_{B^*_0}$ improves previous LSR estimate\,\cite{SNbc15} quoted in Table\,2. It suggests that the experimental candidate $B^*_J(5732)$ can be fairly identified with a $0^{++}$ B-like meson.
The decay constant of ${B^*_0}$ has been already estimated in \,\cite{SNbc15} within LSR. It is quoted in Table\,3 and agrees with the one in\,\cite{WANG}. 
%%%%%%%%%%%%%%%%%%%%%%%%%%%%%%%%%%%%%%%%%
\vspace*{-0.5cm}
\begin{table}[hbt]
{\scriptsize
 \caption{Values of the masses from LSR and HQS compared with lattice  and potential models (PM) results.}}
\setlength{\tabcolsep}{0.35pc}
    {\small
  \begin{tabular}{ccccc}
 % \begin{tabular*}{\textwidth}{@{}l@{\extracolsep{\fill}}llll}
% {\begin{tabular}{@{}llll@{}} \toprule
&\\
\hline
\hline
%\\
Channel&LSR& HQS &Lattice\,\cite{LATT}&PM\,\cite{QUIGG}  \\
%\\
\hline
%\\
Masses \\
$B^*_c(1^{--})$&$6451(86)$  &6315(1)&6331(7)&6330(20)\,\cite{BAGAN}\\
$B^*_{0c}(0^{++})$&6689(198)&6723(29)&6712(19)&6693 \\
$B_{1}(1^{++})$&6794(128)&6730(8)&6736(18)&6731 \\
$B_{c2}(2^{++})$&--&6741(8)&--& 7007\\
%&\\
\hline 
%&\\
$B^*_{0}(0^{++})$&5701(196)\,\cite{SNhl05}&5733& \\

\hline\hline
\end{tabular}
}
\label{tab:hmass}
\end{table}
\vspace*{-0.25cm}
%%%%%%%%%%%%%%%%%%%%%%%%%%%%%%%
\section{Summary and Comparison with some Other Estimates}
%%%%%%%%%%%%%%%%%%%%%%%%%%%%%%%
\subsection*{\b Spectra}
%%%%%%%%%%%%%%%%%%%%%%%%%%%%%%%%
-- We collect the results for the masses obtained  in the previous sections in Table\,2 which we compare with some recent Lattice calculations and Potential Models (PM) results. 

-- We notice that, within the errors, the results from different approaches agree each other except the one for the $2^{++}$ meson. Note that the orginal numbers from\,\cite{QUIGG} are quoted without any errors but we expect that the PM results are known within 20 MeV error as estimated in\,\cite{BAGAN}. 
%%%%%%%%%%%%%%%%%%%%%%%%%%%%%%%%%%%%%%%%
%%%%%%%%%%%%%%%%%%%%%%%%%%%%%%
\begin{table}[hbt]
%\label{tab:param}
{\scriptsize
 \caption{Values of the decay constants $f_H$ in units of MeV using as input the values of the masses from LSR and HQS quoted in Table\,2. } } 
 \vspace*{0.25cm}
\setlength{\tabcolsep}{0.25pc}
    {\small
  \begin{tabular}{ccccc|c}
 % \begin{tabular*}{\textwidth}{@{}l@{\extracolsep{\fill}}llll}
% {\begin{tabular}{@{}llll@{}} \toprule
%&\\
\hline
\hline
%\\
Masses&$B_c(0^{--})$&$B^*_c(1^{--})$ &$B^*_{0c}(0^{++})$&$B_{1c}(1^{++})$&$B^*_{0}(0^{++})$\\
%\\
\hline
LSR&371(17)\,\cite{SNbc20}&442(44)&155(17)&274(23)&-- \\
HQS&--&387(15)&158(9)&266(14)&271(26)\,\cite{SNbc15} \\

\hline\hline
\end{tabular}
}
\label{tab:hdecay}
%\caption{%\scriptsize   
\end{table}
%%%%%%%%%%%%%%%%%%%%%%%%%%%%%%%%%%%%%%%%%%

-- One should mention that the errors from LSR are relatively large which are mainly due to the large range of $t_c$-values. The predictions  can only be improved after a complete measurement of the spectral functions which is out of reach at present. 

-- The quoted errors from HQS come only from the data.  We are aware that some systematic errors not included here are present using the HQS results to LO but the agreement of these results with the accurate data may indicate that these corrections are small. The inclusion of such HQS higher order corrections is beyond the scope of this paper.
%%%%%%%%%%%%%%%%%%%%%%%%%%%%%%%
\vspace*{-0.25cm}
\subsection*{\b Decay constants}
%%%%%%%%%%%%%%%%%%%%%%%%%%%%%%%%
-- The new predictions for the decay constants are collected in Table\,3. One should notice that the values of the decay constants are largely affected by the value of the masses. 

-- We have also  reported in Table\,3, the predictions using the relatively precise predictions on the masses from HQS. 

-- The difference of the LSR and HQS results for $f_H$ is due to the LO factor Exp$[M^2_H\tau/2]/M_H^2$ entering in the LSR expression of $f_H$. 

%%%%%%%%%%%%%%%%%%%%%
%\section*{References}
%%%%%%%%%%%%%%%

 %%%%%%%%%%%%%%%%%%%%%%%%%%%%%%%%%%%
 \end{document}